\documentclass[%
 aip,
 jcp,%
 amsmath,amssymb,
]{revtex4-1}

\usepackage{graphicx}% Include figure files
\usepackage{dcolumn}% Align table columns on decimal point
\usepackage{bm}% bold math
\usepackage{multirow}
\usepackage{color}
\usepackage{cancel}
\begin{document}

%\preprint{AIP/123-QED}

\title{A New Approach For Learning Coarse-Grained Potentials with Application to Immiscible Fluids}
%\thanks{Footnote to title of article.}

\author{Peiyuan Gao}
\email{peiyuan.gao@pnnl.gov} 
\author{Xiu Yang} 
\author{Alexandre M. Tartakovsky}
\email{alexandre.tartakovsky@pnnl.gov}
\affiliation{Pacific Northwest National Laboratory, Richland, Washington
99352, USA}
\date{\today}% It is always \today, today,

\begin{abstract}
 Even though atomistic and coarse-grained (CG) models have been used to simulate liquid nanodroplets in vapor,
very few rigorous studies of the liquid-liquid interface structure are available, and most of them are limited to planar interfaces.
In this work, we evaluate several existing force fields (FF)s, including two atomistic and three CG FFs,
with respect to modeling the interface structure and thermodynamic properties of
the water-hexane interface. Both atomistic FFs are able to quantitatively
reproduce the interfacial tension and the coexisting
densities of the experimentally-observed planar interface. We use the atomistic FFs to model water droplets in hexane and use these simulations to test the CG FFs. We find that the tested CG FFs cannot
reproduce the interfacial tensions of planar and/or curved interfaces.
Finally, we propose a new approach for learning CG potentials within the CG SDK (Shinoda-DeVane-Klein) FF framework from atomistic simulation data. We demonstrate that the new potential significantly improves the prediction of both the interfacial tension and structure
of water-hexane planar and curved interfaces. 
\end{abstract}

\maketitle

\section{Introduction}
Many physical, chemical, and biological processes such as micelle formation,
interfacial polymerization, and protein folding occur in presence of  
hydrophobic-hydrophilic interfaces.\cite{RN1,RN2,RN3} Therefore,
understanding the interface at the molecular level is fundamentally
important. 
%The linking of macroscopic properties such as interfacial
%reactivity, solubility, and permeability and the microscopic interfacial
%properties such as structure and local interfacial tension, which is difficult to measure 
%experimentaly, recently attracted more attention in theory and simulation studies.\cite{RN4}  
In contrast
with liquid-vapor interfaces, simulation of a liquid-liquid interface is
more challenging because a large system size is required to stabilize the
liquid phases and the interfacial region.\cite{RN30} The
computation of interfacial properties of large systems involving 
sampling of long time and large length scales remains a challenge for
atomistic models.\cite{RN31,RN32,RN33} Therefore, coarse-grained (CG) models present an
attractive alternative to atomistic models because of their ability 
to cover much larger time and length scales.\cite{RN34} To this end, several methods to obtain CG force fields have been developed by
averaging atomistic details to reproduce certain 
essential properties of complex fluids. For example, the SDK (Shinoda-DeVane-Klein) CG force field (FF)
was shown to accurately model the surface tension, bulk density, and
hydration free energy of water and alkanes.\cite{RN23} The MARTINI FF,
originally designed for lipids, surfactants, and biomacromolecules, was
also used to model the liquid-vapor interface of
an alkane.\cite{RN25,RN26} 
The SAFT (Statistical Associating Fluid
Theory) CG FF\cite{RN27,RN29} was developed for many solvents, including  
water, alkanes, and carbon dioxide, where the effective CG
intermolecular interactions between particles are estimated using an
accurate description of the macroscopic experimental vapor-liquid
equilibria data by means of a molecular-based equation of state. 
The
above-mentioned CG FFs were shown to accurately describe
multiple physical properties for common industrial fluids. 
However, even for relatively simple binary multiphase flow system, the accuracy of the CG FF
interface structure and physical properties predictions has not been well studied.

Machine learning methods have been successfully used to construct
 a potential energy response surface using quantum chemistry calculations and parameterize atomistic potentials.\cite{RN781, RN783,RN785,RN784} 
In this paper, we use a machine learning method to parameterize a CG FF for a water-hexane system 
 and compare it with existing CG FFs. We select the water-hexane system as 
a typical immiscible binary system.  
We investigated the CG FFs ability to
accurately model the surface tension and intrinsic/non-intrinsic densities of planar and curved interfaces as compared with those observed in experiments and/or atomistic simulations.
This paper is organized as
follows. Section \ref{Sim_methods} describes the atomic and CG models. Section \ref{results} discusses the atomic and CG simulation results. Section \ref{ML_methods} introduces the machine learning method and discusses its results. 
Section \ref{conclusions} 
presents the conclusions and outlook for CG modeling of complex liquid-liquid
interfaces.

\section{Simulation models and methods}\label{Sim_methods}

\subsection{Atomistic model and simulation}

Several water models have been proposed in literature, but only the TIP4P2005 (Transferable
Intermolecular Potential with 4 Points (2005)) model was shown to accurately
reproduce the temperature-dependent liquid-vapor surface tension.\cite{RN36,RN37}
Therefore, we select the TIP4P2005 water model in our atomistic simulations.
The TraPPE
(Transferable Potentials for Phase Equilibria)
FF\cite{RN38} was shown to 
predict surface tension of alkanes in experiments.\cite{RN38}Also, Neyt \emph{et al.} demonstrated that the
 TIP4P2005 water and octane models combination in TraPPE FF can
reproduce the experimentally measured interfacial tension of a water/\emph{n}-octane system.\cite{RN39} Therefore, in this work we also choose the \emph{n}-hexane
model from the TraPPE FF. The interaction potential 
between the TIP4P2005-modeled water  and the TraPPE-modeled alkane is modified
following Ashbaugh's protocol.\cite{RN40} This modification
improves the hydration energy of an alkane molecule in water and does
not change other properties. In addition, note that the \emph{n}-hexane model in TraPPE FF is  united-atom model, where CH\textsubscript{3} and
CH\textsubscript{2} groups are represented with a single united atom. Therefore, the interaction potential between the ``TIP4P2005'' water and TraPPE \emph{n}-hexane does not include electrostatic interactions, which might affect the interfacial structure and tension. 
To study the effect of this potential on
interfacial tension, we also test the hexane model in the Optimized Potential for
Liquid Simulation All-Atom\cite{RN41} (OPLS-AA) FF.  

In our simulations of planar interfaces, we put a pre-equilibrated water slab sandwiched between pre-equilibrated \emph{n}-hexane slabs. The initial simulation box size is  $L_x=L_y=6$ nm and $L_z = 20$ nm. 
We place 8315 water molecules and 1152 hexane molecules for the TraPPE FF and 1140 hexane molecules for the OPLS-AA FF in the simulation box. 
Initially, the water and hexane molecules are separated by the plane interface.  
We also model a spherical water droplet in \emph{n}-hexane with both the TIP4P2005 water-TraPPE \emph{n}-hexane and the TIP4P2005 water-OPLS \emph{n}-hexane models.
We simulate droplets with radii of 2 nm (1026 water
molecules) and 3 nm (3609 water molecules) in the simulation box with $L_x=L_y=L_z\approx11$ nm and $L_x=L_y=L_z\approx15$ nm, respectively. %Then we add \emph{n}-hexane molecules into the simulation box. 
The box size is slightly adjusted during the equilibration process to keep pressure at 1 atm.    
For both the curved and planar interfaces, the long-range dispersion
force correction method is used to obtain the correct density and pressure. 
These planar and droplet systems are equilibrated using the NP\textsubscript{N}AT \cite{RN724} ensemble (to keep the pressure constant, the box volume 
is changed by varying $L_z$) and the NPT ensemble with V-rescale
thermostat and Berendsen barostat for 10 ns, respectively. The temperature and pressure are set to 310 K and 1 atm. Then we run another 10 ns simulation with the canonical ensemble at 310 K to collect data. All bonds 
between atoms are fixed by the LINCS algorithm.\cite{RN42} Periodic
boundary conditions are used in all three directions. The time step
is 2 fs. All the atomistic simulations are performed with GROMACS.  

\subsection{CG model and simulation}
We selected the MARTINI (including the original and polarized water model) and SAFT CG FFs for modeling the water-hexane
interface. 
For the original MARTINI water model we replaced the 12\% CG water beads with anti-freeze CG water beads \cite{RN25} to prevent the water from freezing. Our simulation results show
that the addition of anti-freeze CG water beads does not affect the
interfacial tension between water and  \emph{n}-hexane until 50\% of the total number of water beads.
For polarized MARTINI water model, the anti-freezing CG water bead is not needed. We select the bio2 CG water model in the SAFT CG FF.

We build planar and curved interface systems for
all considered CG FFs. In the planar interface simulations, the
simulation box size is set to \(L_{x} = L_{y} > 5\sigma\) and
\(L_{z} > 11\sigma\) to avoid the boundary effect on the surface
tension.\cite{RN53} To study properties of curved interface, we simulate a 2 nm water
droplet in \emph{n}-hexane. 
%We use cubic shaped boxes for the curved systems. 
To reduce the boundary effect, the initial
length of the simulation box is set to 11 nm. The simulation boxes are
equilibrated for 20 ns in NP\textsubscript{N}AT and NPT ensembles at 310 K and 1 atm, respectively. Then,
we perform 30 ns (planar interface) and 10 ns (curved interface) NVT 
simulation at 310 K for data collection. To get better statistics, we performed 
five parallel simulations for each curved interface system. 
The cutoffs for vdW interaction are 1.2 and
1.5 nm for the MARTINI and SAFT CG FFs, respectively. The cutoff for
Coulomb potential is 1.2 nm for the polarized water model in the MARTINI CG
FF. The V-rescale thermostat and Berendsen barostat are used to keep
constant temperature and pressure during pre-equilibrium. 
The Nose-Hoover thermostat is then employed in the production simulation.
The time step is 10 fs. All CG simulations are performed with GROMACS.

\section{Simulation results}\label{results}
In this section, we investigate the water and hexane density and pressure profiles and interfacial tensions of water-hexane systems with planar and curved interfaces using two atomistic and three CG FFs. 
%A comparison is also made between atomistic and CG simulations. 
%
Our analysis demonstrates that the three considered CG FFs cannot reproduce the interfacial tension of the both planar and curved interfaces observed in the atomistic simulations. In Section \ref{ML_methods}, we develop a new CG interfacial potential using a machine learning approach.     

\subsection{Planar interface}

Here, we model a slab of water sandwiched between two slabs of hexane forming two planar water-hexane interfaces approximately  located at $z$=7 and 13 nm. In Section \ref{density}, we present averaged density profiles as a function of the normal distance from the interfaces. In Section \ref{pressure}, we show pressure profiles as a function of $z$. Pressure and density profiles are averaged over the $x$ and $y$ coordinates and time.

\subsubsection{Density profiles}\label{density}

\begin{figure*}
\includegraphics[scale=0.35]{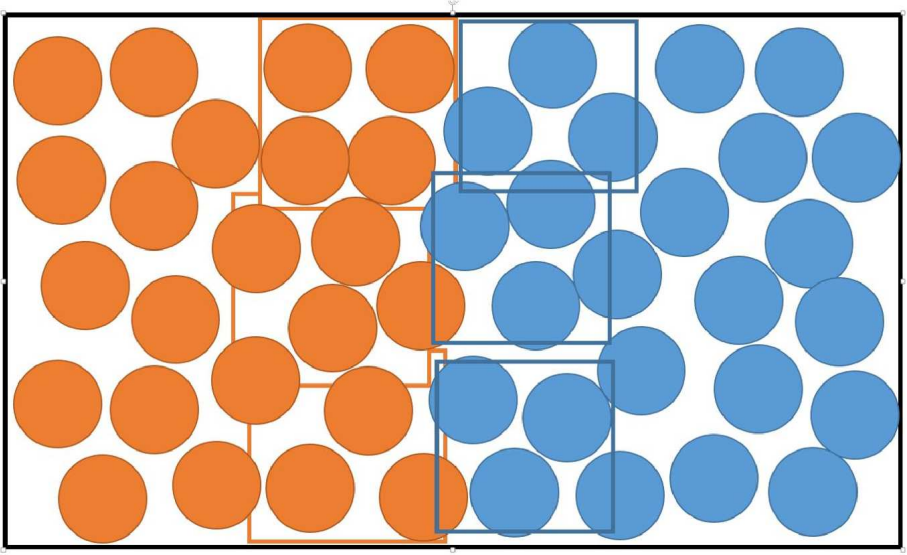}
\caption{The scheme of the intrinsic density calculation for a water-hexane interface.}
\label{fig:scheme}
\end{figure*}

\begin{figure*}
\includegraphics[scale=0.35]{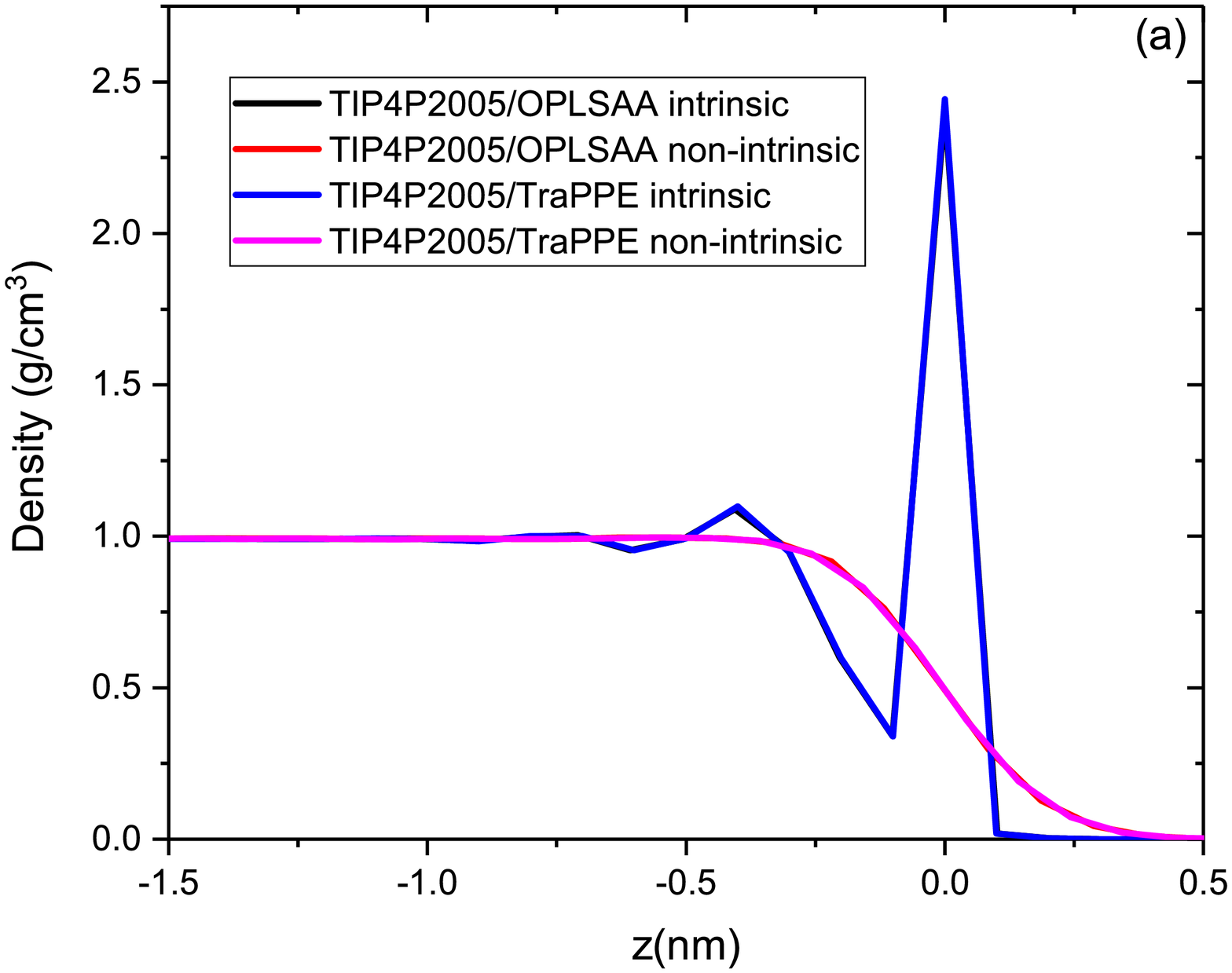}
\includegraphics[scale=0.35]{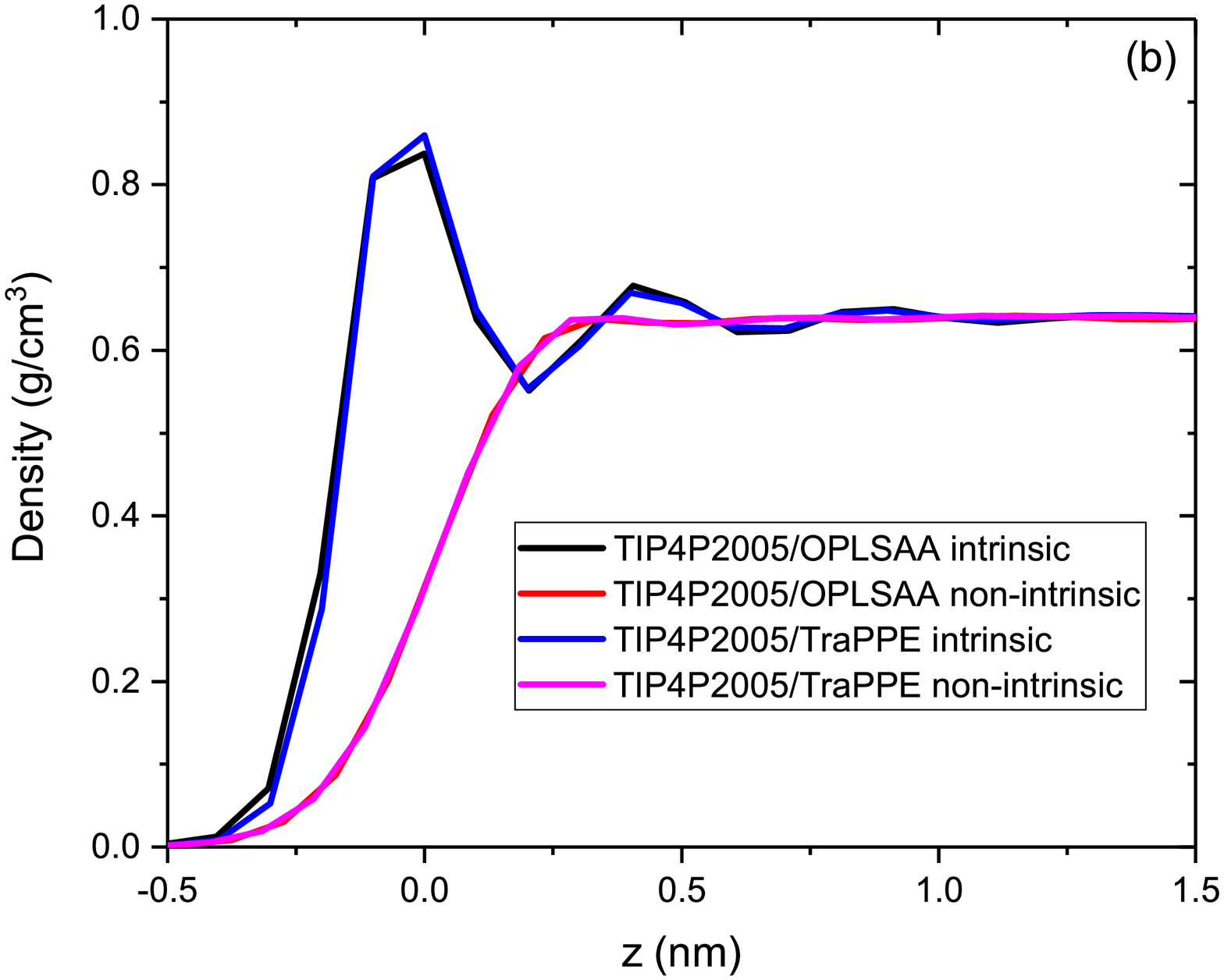}
\caption{The intrinsic and non-intrinsic 
density profiles of (a) water and (b) hexane at the water-hexane interface as a function of $z$ obtained from atomistic simulation. The point $z=0$ corresponds to the
position of the outermost water/hexane atoms in the intrinsic density
profile and the Gibbs dividing surface of the water-hexane system for the
non-intrinsic density profile.}
\label{fig:AA_density}
\end{figure*}

We use intrinsic and non-intrinsic densities of liquids to describe the studied water-hexane systems. 
The non-intrinsic or local mass density $\rho_N(x)$ is defined as the mass of liquid in a cube centered at fixed locations $x$ per volume of the cube. The intrinsic density $\rho(r)$ is computed as the mass of liquid in a cube centered at locations $r$ that move with the interface (see Figure \ref{fig:scheme} and Appendix \ref{App_densities} for details about computing the densities). The intrinsic density provides more information about the interface structure (i.e., the location of the interface and the molecular organization) than the non-intrinsic density.\cite{RN612} The non-intrinsic density profile is smooth and only contains approximate information about the interface location. The intrinsic density profile has local peaks corresponding to the locations of molecules layers near the interface, with the largest peak corresponding to the location of the interface.\cite{RN54,RN55,RN56,RN57}
Figure \ref{fig:AA_density} presents the intrinsic and non-intrinsic density profiles 
of
water and \emph{n}-hexane of a water-hexane planar interface obtained from atomistic simulations with the TIP4P2005-TraPPE and TIP4P2005-OPLSAA models. Both atomistic models result in the same water density profiles and very similar hexane density profiles. Also, both atomistic models can
reproduce the experimental density of water and hexane at 310 K. The intrinsic density profiles show that there are two water layers close to the interface. 
%Furthermore, the water intrinsic density profile across the interface does not depend on the atomic model of n-hexane model. This is consistent with previous MD simulation
%results with different water model.\cite{RN56} 
In addition, the strong directional bonding of water creates a well-defined correlation structure at short
distances from the interface, but it does not propagate to longer distances as efficiently
as it does for more packed liquid structures such as alkanes. The comparison of Figure \ref{fig:AA_density} (a) and (b)
shows longer-range oscillations in alkanes than in water. Similar observations were made for a water-hexane binary
system with the SPC/E water model.\cite{RN72} In the case of hexane, we see that the distribution of the first peak is wider. This is due to the long tail of the alkane molecule.
Overall, we find that the intrinsic
structure of the water/\emph{n}-hexane system is insensitive to 
atomistic FF parameters.

\begin{figure*}
\includegraphics[scale=0.35]{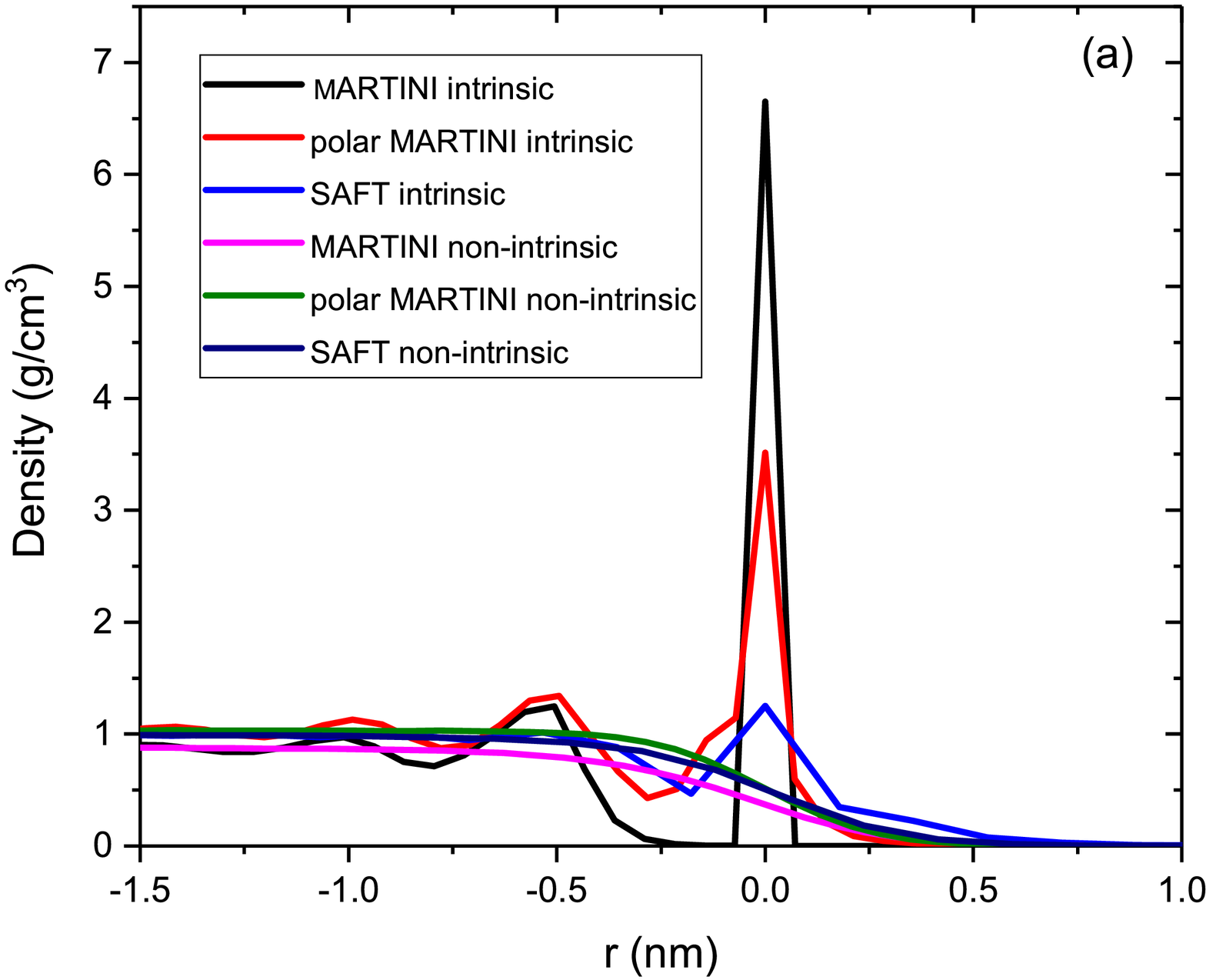}
\includegraphics[scale=0.35]{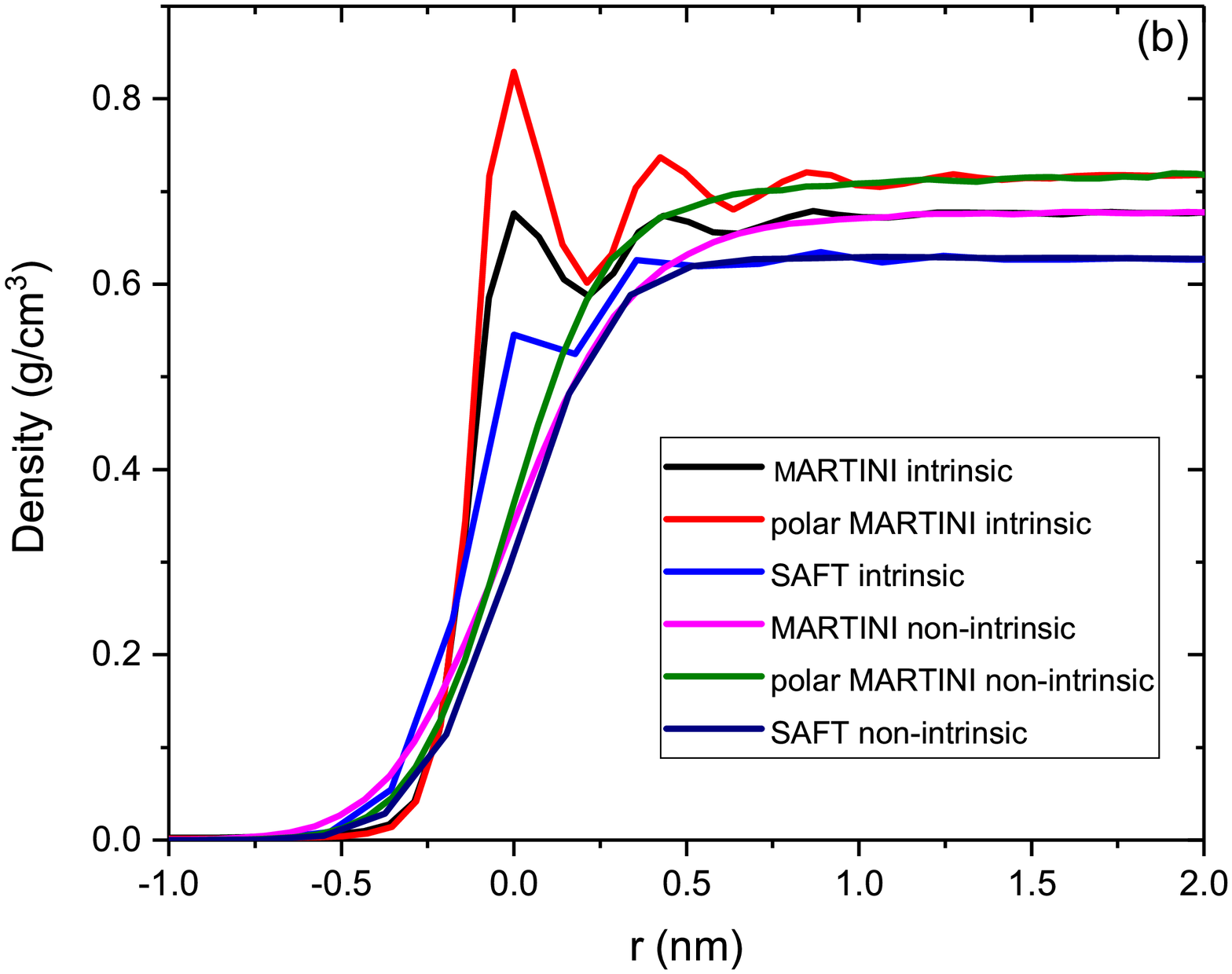}
\caption{The intrinsic and non-intrinsic 
density profiles of (a) Water and  (b) Hexane at the water-hexane interface in CG simulations. 
 The zero point of the interface
corresponds to the position of the outermost water/hexane atoms in the
intrinsic density profile and the Gibbs dividing surface of the water-hexane
system for the non-intrinsic density profile.}
\label{fig:CG_density}
\end{figure*}

The density profiles in CG simulations are shown in Figure \ref{fig:CG_density}. The non-intrinsic and intrinsic densities of water and hexane are different for various CG FFs. The non-intrinsic density profile obtained with the SAFT CG FF is flatter than the MARTINI CG FF. 
It should be noted that the density of water in the \emph{n}-hexane
phase in the atomistic simulations is negligibly small for the  considered here TIP4P2005 water model  (0 due to the float point precision in the atomistic simulations). 
In CG simulations, the density of water in hexane is  3$\times$10\textsuperscript{-4}
g/cm\textsuperscript{3} for the MARTINI CG FF that is approximately five times larger than the experimental value of 6$\times$10\textsuperscript{-5} g/cm\textsuperscript{3}.\cite{RN956} For the SAFT bio2 CG water model in \emph{n}-hexane, the water density is even greater.
In Figure \ref{fig:CG_density}(a), the intrinsic water density profile has three  peaks (two peaks were observed in MD simulations). This indicates that the CG water phase shows a longer-range ordered structure in comparison with the atomistic simulation. The intrinsic density
profiles are similar for the original and polarized MARTINI CG water models,
except that the original MARTINI CG water model has a higher interfacial density.
The first peak in 
the SAFT bio2 CG model is lower than in the atomistic models because the CG model produces a wider interface. The positions of the first intrinsic density peaks for CG \emph{n}-hexane models
are also very close. 
The hexane intrinsic density profiles, obtained from the MARTINI and SAFT CG
\emph{n}-hexane models,  do not have distinct peaks (Figure \ref{fig:CG_density}(b)). 
However, we can observe a peak in the hexane intrinsic density in the polarized MARTINI CG model. 
This is because single CG beads are used for both the MARTINI and SAFT bio2 CG water models.
On the other hand, the polarized
MARTINI CG water model has a physics-based three-point structure. 
Bresme \emph{et al.} have demonstrated that the packing of
water molecules will influence the orientation of alkane molecules at
the interface.\cite{RN55} In our CG simulations, we also see that
the geometry topology constraint of the CG water model would affect the local
surface structure of the hexane phase. For the SAFT \emph{n}-hexane model, the CG water beads infiltrate
into the hexane phase so deeply that the density of the first peak is lower than that of the
bulk phase. In Figures \ref{fig:CG_density} (a) and (b), the intrinsic density
profiles of water or hexane are all different, which illustrates that the intrinsic density
profile is sensitive to the choice of water and \emph{n}-hexane CG models. 
%This behavior is not the same as the aforementioned atomistic
%models.

\subsubsection{Pressure profiles and interfacial tension}\label{pressure}

\begin{figure}
\includegraphics[scale=0.35]{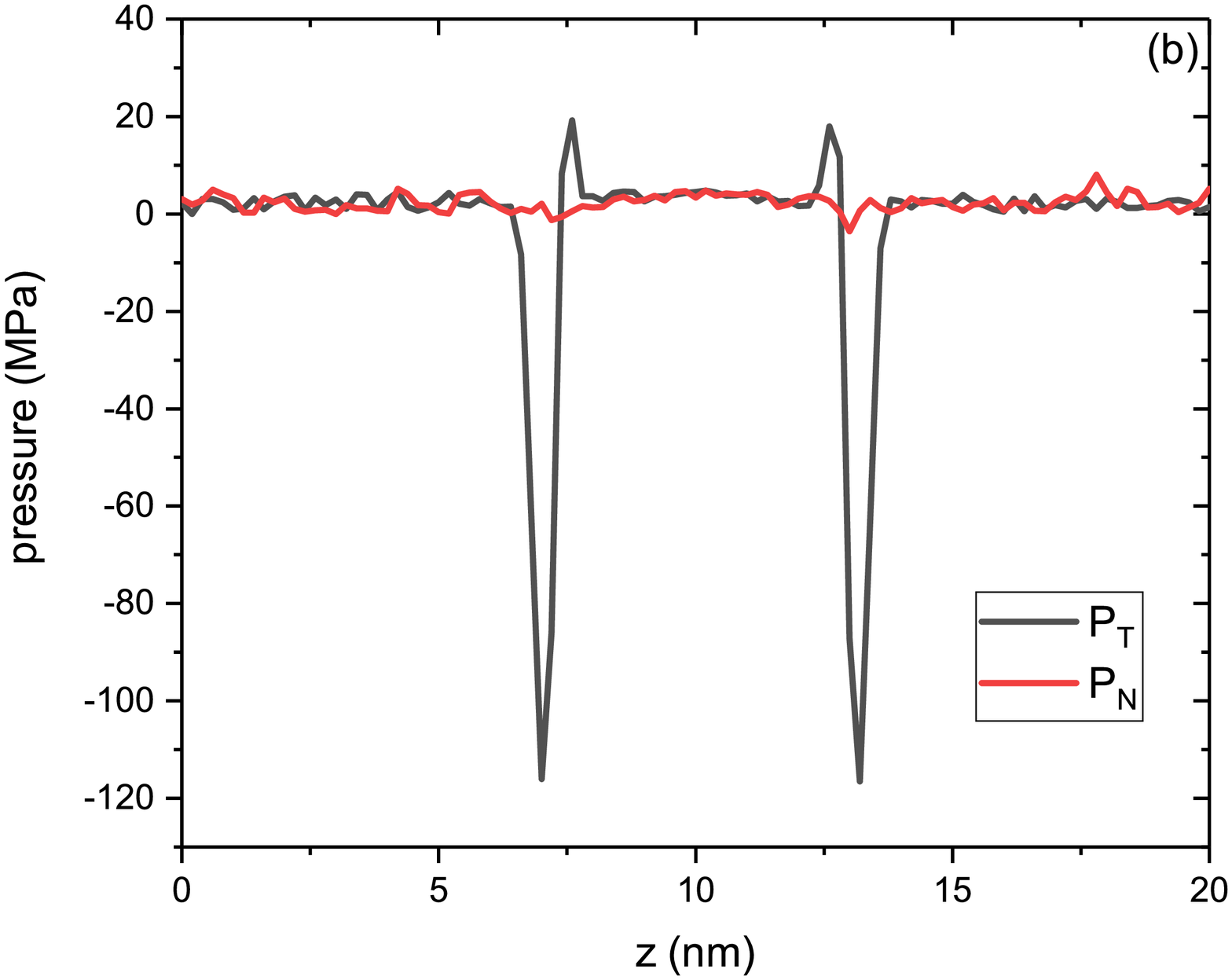}
\includegraphics[scale=0.35]{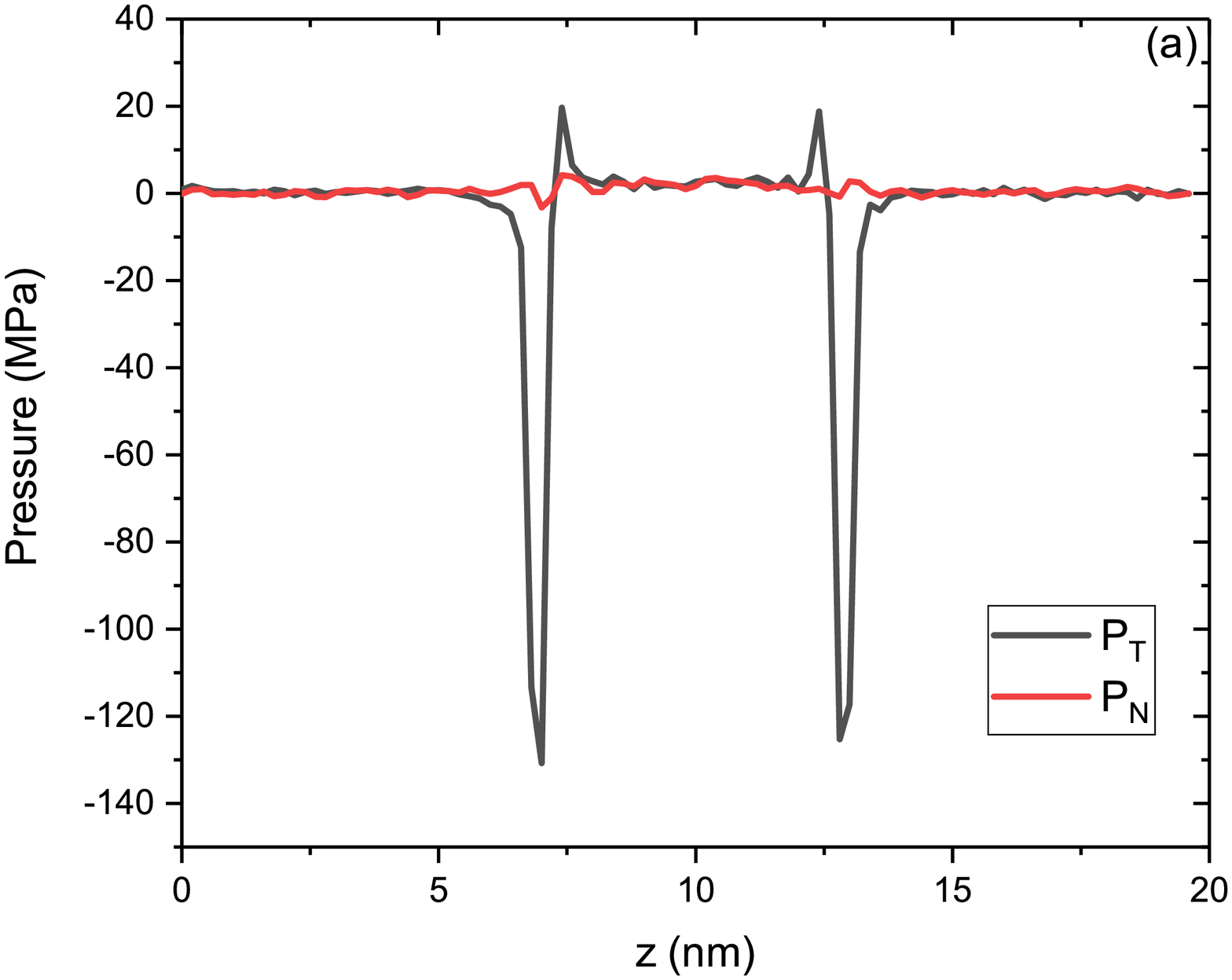}
\caption{Pressure tensor components of the water-hexane planar interface in the atomistic 
(a) TIP4P2005 water + hexane in
TraPPE FF and (b) TIP4P2005 water + hexane in OPLS-AA FF models.}
\label{fig:AA_pressure}
\end{figure}

\begin{figure}
\includegraphics[scale=0.35]{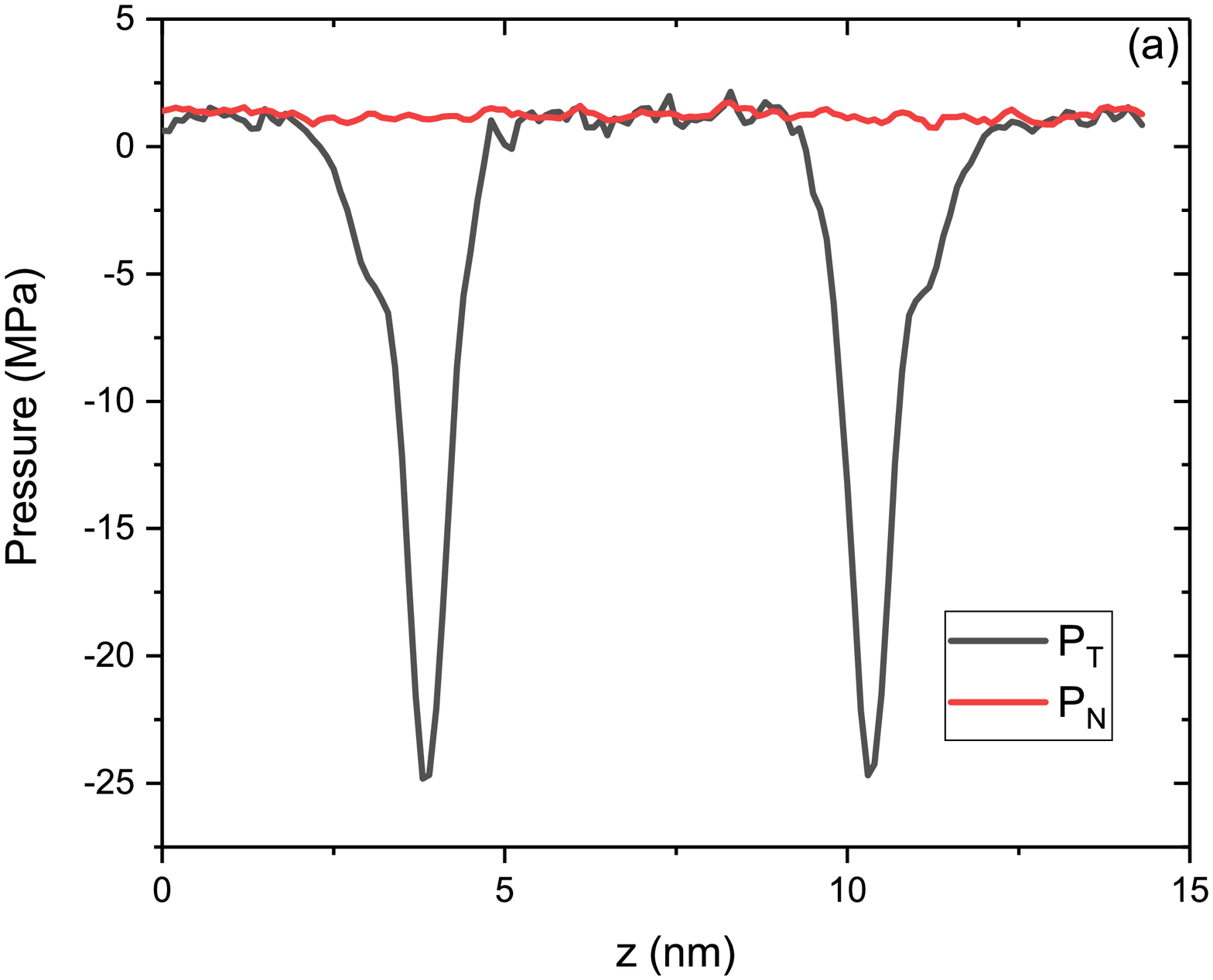}
\includegraphics[scale=0.35]{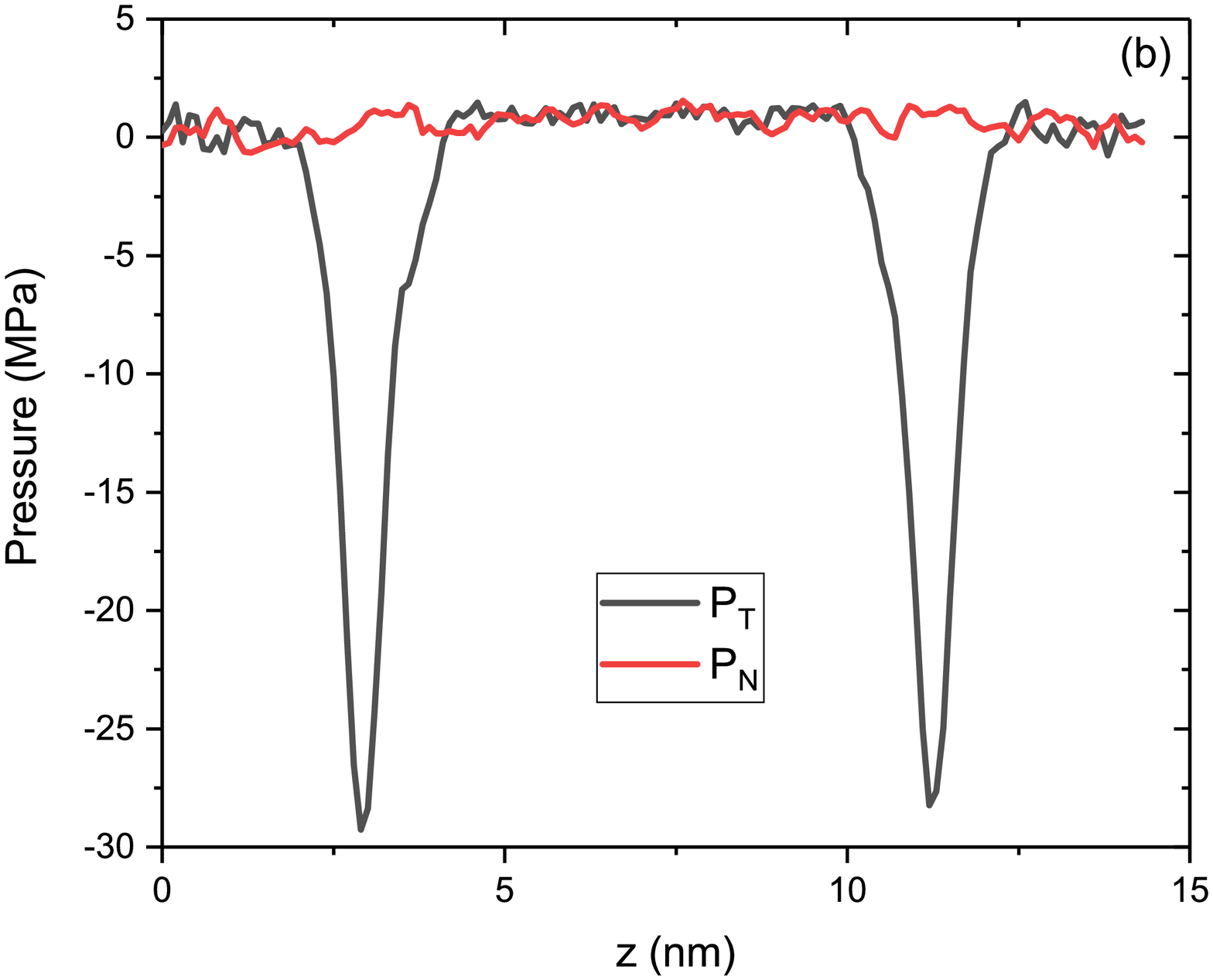}
\includegraphics[scale=0.35]{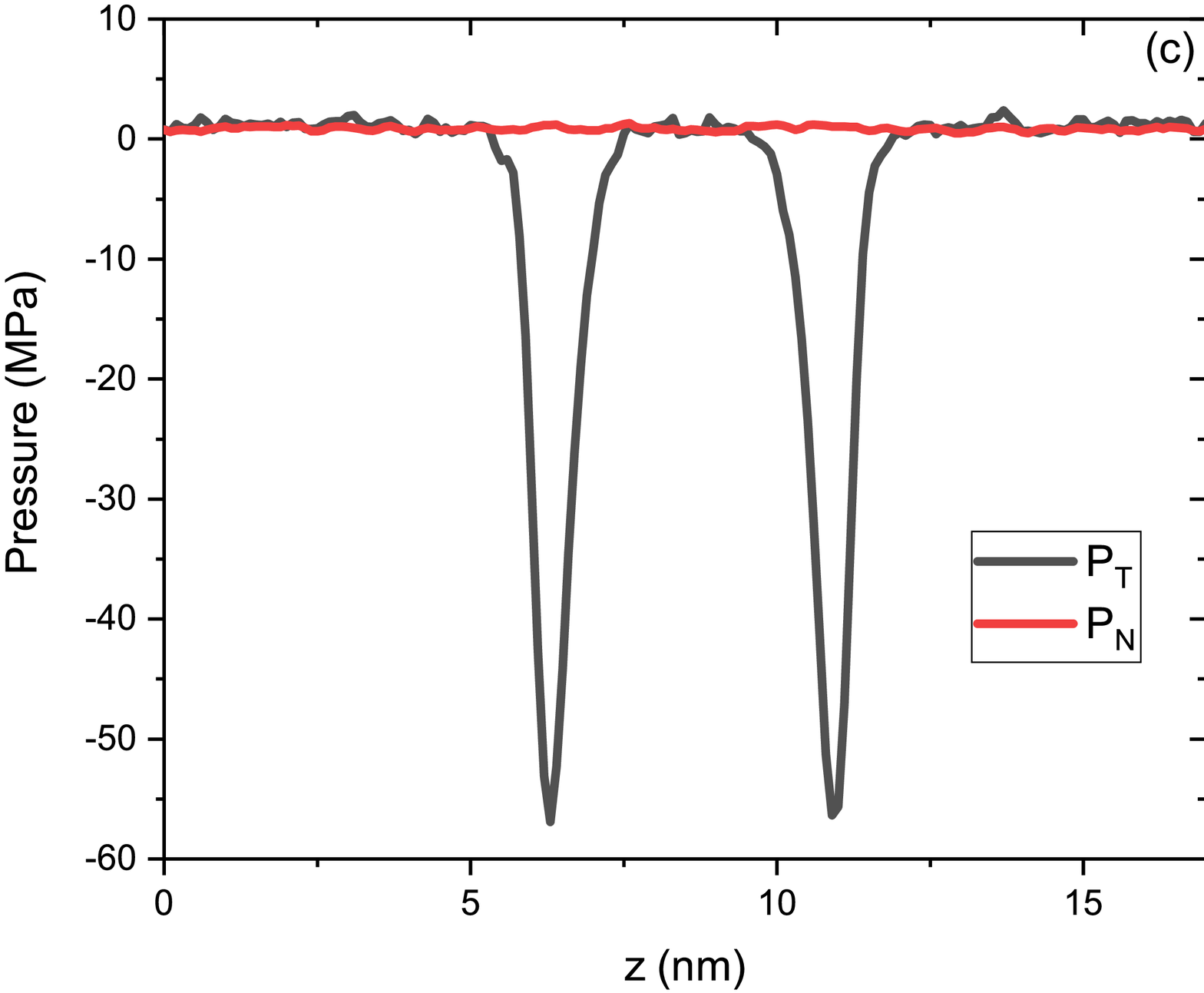}
\caption{Local pressure components of water-hexane planar interface in 
 (a) MARTINI FF, (b) MARTINI FF with polarized
water, and  (c) SAFT FF models.}
\label{fig:CG_pressure}
\end{figure}

Previous atomistic simulations demonstrated  that the errors in the estimated surface tension and liquid density are closely correlated.\cite{RN73,RN36} Therefore, the
accurate prediction of density is very important in the calculation of surface
tension. Above, we demonstrated that both considered atomistic water and hexane models can reproduce the
liquid bulk density at 310 K. Here, we calculate the interfacial tension 
using the pressure tensor based on the mechanical approach.\cite{RN62,RN63,RN66,RN67,RN69} The details off calculating the pressure tensor components and the interfacial tension are given in Appendix \ref{App_pressure}. The normal and tangent pressure tensor 
components as a function $z$, obtained from the atomistic and CG simulations, are presented in Figure \ref{fig:AA_pressure}
and Figure \ref{fig:CG_pressure}. There are two symmetrical positive stress regions
in the tangent component of pressure, corresponding to the two water-hexane interfaces, in the atomistic simulations (Figure \ref{fig:AA_pressure}). They both
appear on the water side of the interfaces. A similar pressure profile was also
observed in an atomistic simulation with the TIP3P water and CHARMM hexane
models.\cite{RN74} Water molecules cause interface polarization and the positive
pressure region on the water side of the interface.

Calculated and experimentally determined interfacial tensions are listed in Table \ref{tableAAIT}. Both atomistic
models predict the interfacial
tension within 5\% of the experimental value. We note that the computational cost of the all-atom model (OPLS-AA FF) is about five times larger
than that of the united-atom model (TraPPE FF). The SAFT CG FF can also reproduce the
experimental interfacial tension. However, the interfacial tension
predicted by the MARTINI CG FF is only half of the experiment value. 
%
%This is similar to Neyt`s result.\cite{RN39} 
%
A previous MARTINI CG FF simulation study of a water-octane system at 298 K also reported an approximately 25\% error in the estimated interfacial tension.\cite{RN39} 
 %Considering our simulation temperature is
%310 K, it further shows that the interfacial tension predicted by
%MARTINI CG FF is sensitive to the temperature, although the temperature
%effect is weak for bulk density of each phase in the binary system.
In addition, we find that using the polarized MARTINI water model instead of the MARTINI water model 
only slightly improves the interfacial tension prediction.

\begin{table}
\caption{The interfacial tensions $\gamma_0$ of the water-hexane planar interface in 
the atomistic and CG simulations and the experiment\cite{RN922}  at 310 K.}
\begin{tabular}{cc}
\hline
Model & Interfacial tension (mN/m) \\
\hline
Experiment & 49.4\\
atomistic TIP4P2005+TraPPE  & 52.4$\pm$1.1\\
atomistic TIP4P2005+OPLS-AA  & 52.1$\pm$1.2\\
CG MARTINI & 25.9$\pm$1.0\\
CG Polarized MARTINI &
27.8$\pm$1.2\\
CG SAFT & 51.6$\pm$1.1\\
\hline
\label{tableAAIT}
\end{tabular}
\end{table}

\subsection{Curved interface}

Here, we model a 2 nm water droplet in hexane.  In Section \ref{droplet_density}, we present averaged density profiles as a function of the normal distance from the interface. In Section \ref{droplet_pressure}, we show averaged pressure profiles as a function of the distance from the droplet center. %Pressure and density profiles are are also averaged over time.    

\subsubsection{Density profiles}\label{droplet_density}

\begin{figure}
\includegraphics[scale=0.4]{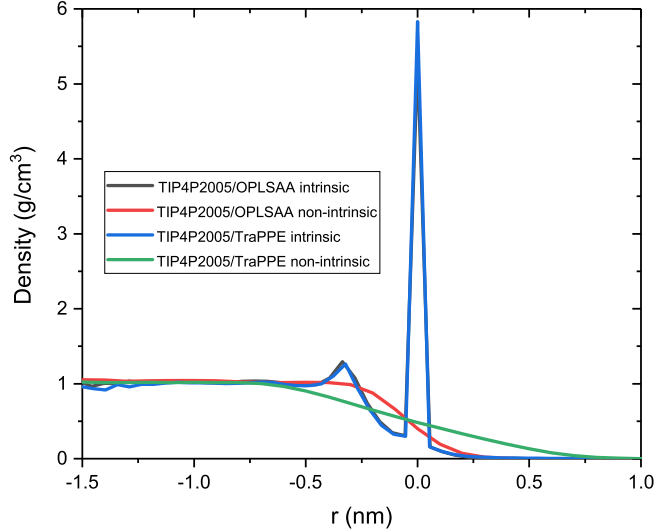}
\caption{The intrinsic and non-intrinsic 
density profiles of a 2 nm water droplet in \emph{n}-hexane obtained from atomistic simulations. }
\label{figAA2nmdensity}
\end{figure}

The intrinsic and non-intrinsic density profiles of a 2 nm water droplet
in \emph{n}-hexane, obtained in the two atomistic models, are shown in Figure \ref{figAA2nmdensity}. There
are two peaks in the intrinsic density profiles in both atomistic models, which is similar to what we observed the planar
interface atomistic simulations. However, the peaks at the curved interface are higher than those at the planar interface.
Compared to Figure \ref{fig:AA_density}, we also see that the width of the first peak is
narrower, implying that the first water layer on the droplet surface
is thinner than the one at the planar interface.

\begin{figure}
\includegraphics[scale=0.4]{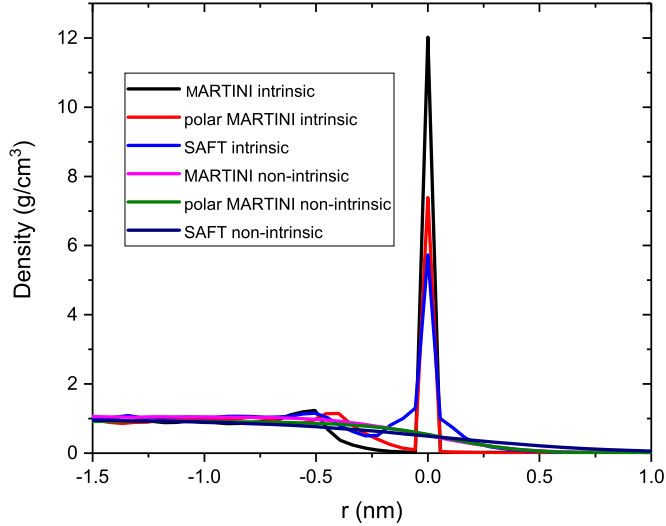}
\caption{The intrinsic and non-intrinsic 
density profiles of a 2 nm water droplet in \emph{n}-hexane obtained from CG simulations.}
\label{figCG2nmdensity}
\end{figure}

The CG water droplets show qualitatively different results. Figure \ref{figCG2nmdensity} shows the
density profiles of a 2 nm water droplet in \emph{n}-hexane with various CG
FFs. In the CG simulations of the planar interface, we see three density
peaks on the water side. In Figure \ref{figCG2nmdensity}, the intrinsic density profile in the SAFT CG FF simulation has 
three peaks, while there are only two peaks for the MARTINI FF. This
could be caused by a larger cutoff in the SAFT CG FF.
Both, the CG and atomistic simulations show that the first peak in the water density profile is much higher for a curved interface than a planar interface. 

\subsubsection{Pressure profiles and interfacial tension}\label{droplet_pressure}

\begin{figure}
\includegraphics[scale=0.35]{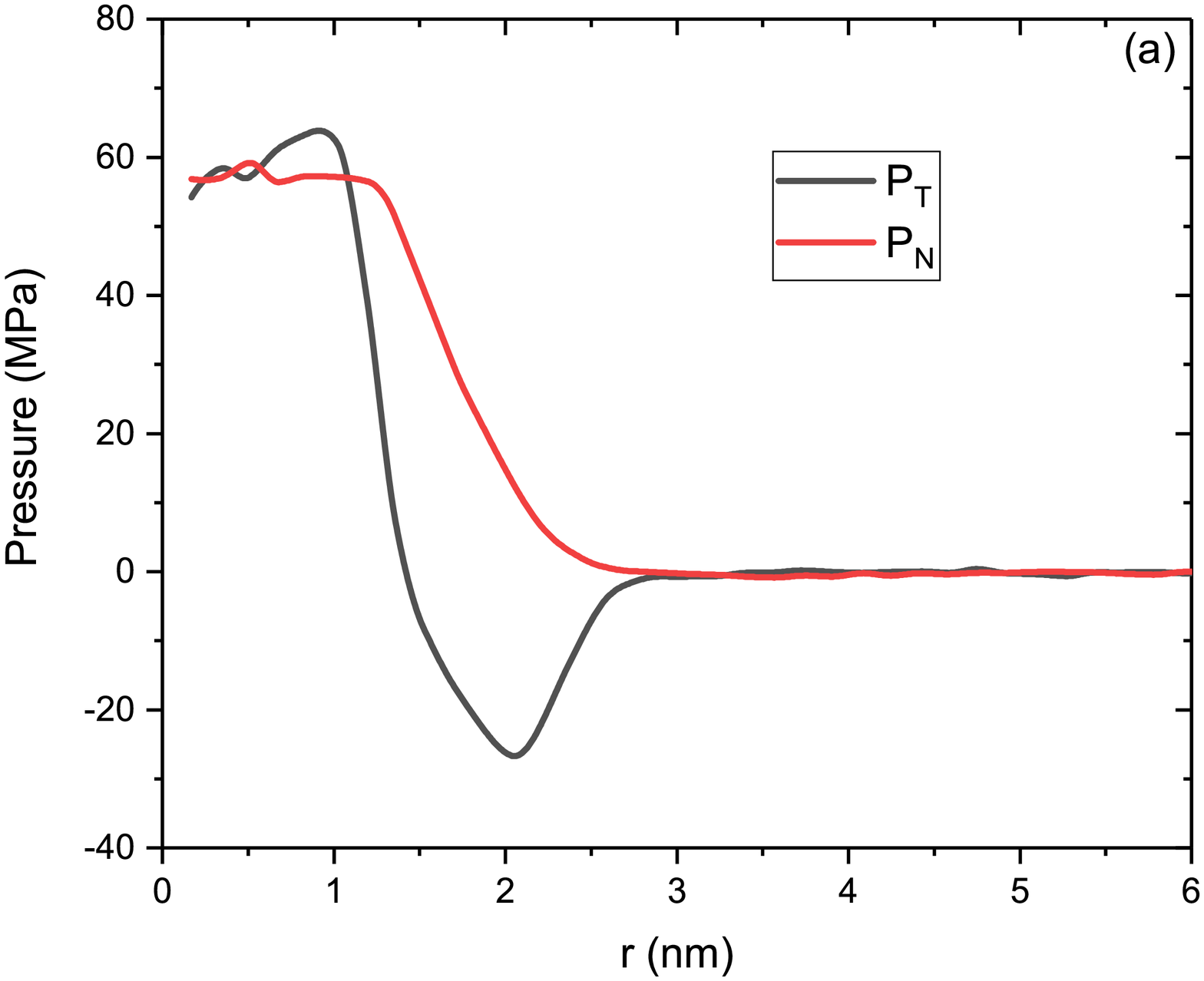}
\includegraphics[scale=0.35]{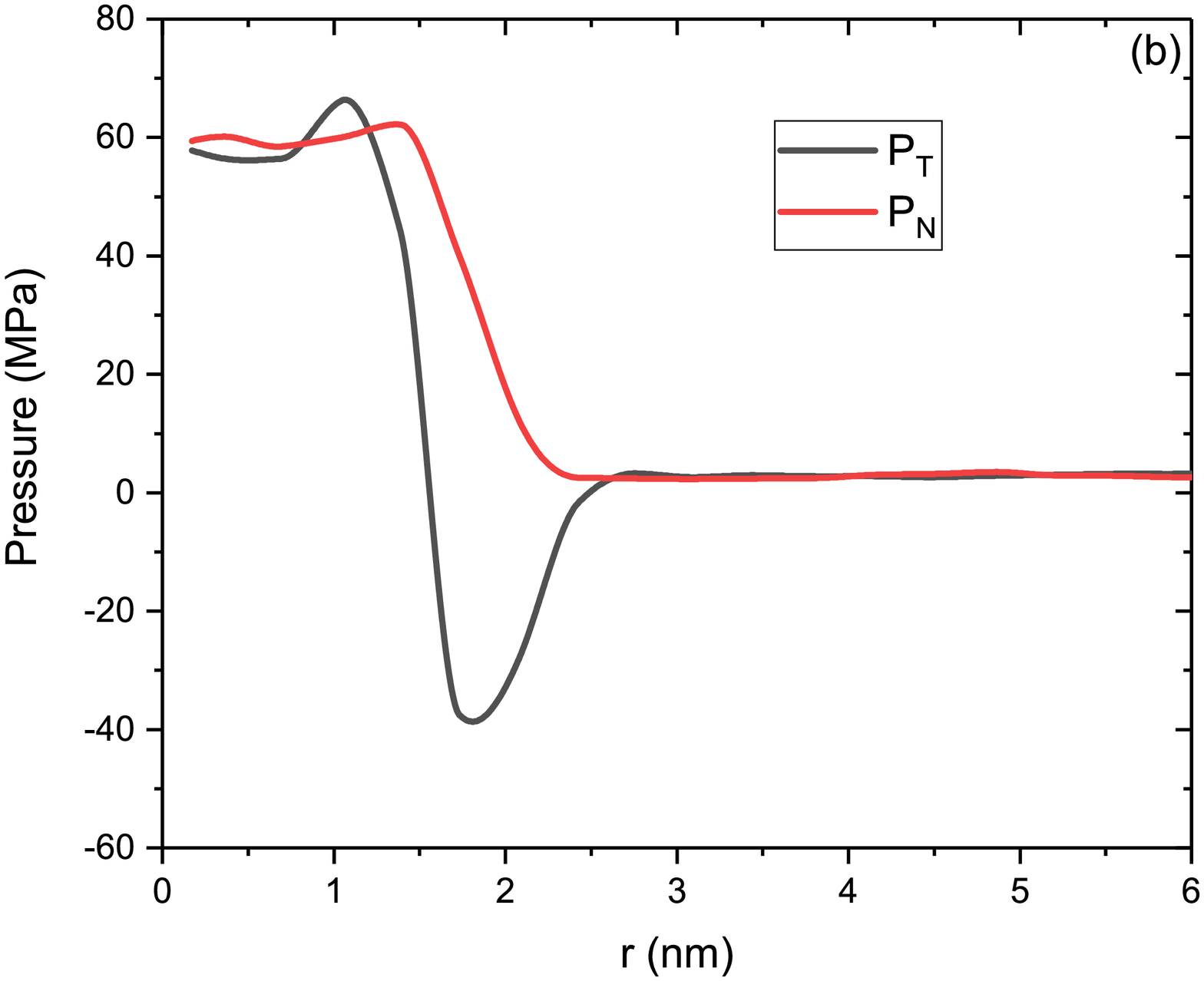}
\caption{Pressure components as a function of the distance from the center of a 2 nm water droplet in \emph{n}-hexane in atomistic FF simulations, including 
 (a) TIP4P2005 water model and \emph{n}-hexane in TraPPE FF and (b)
TIP4P2005 water model and \emph{n}-hexane in OPLS-AA FF.}
\label{fig:AA2nmpressure}
\end{figure}

\begin{figure}
\includegraphics[scale=0.35]{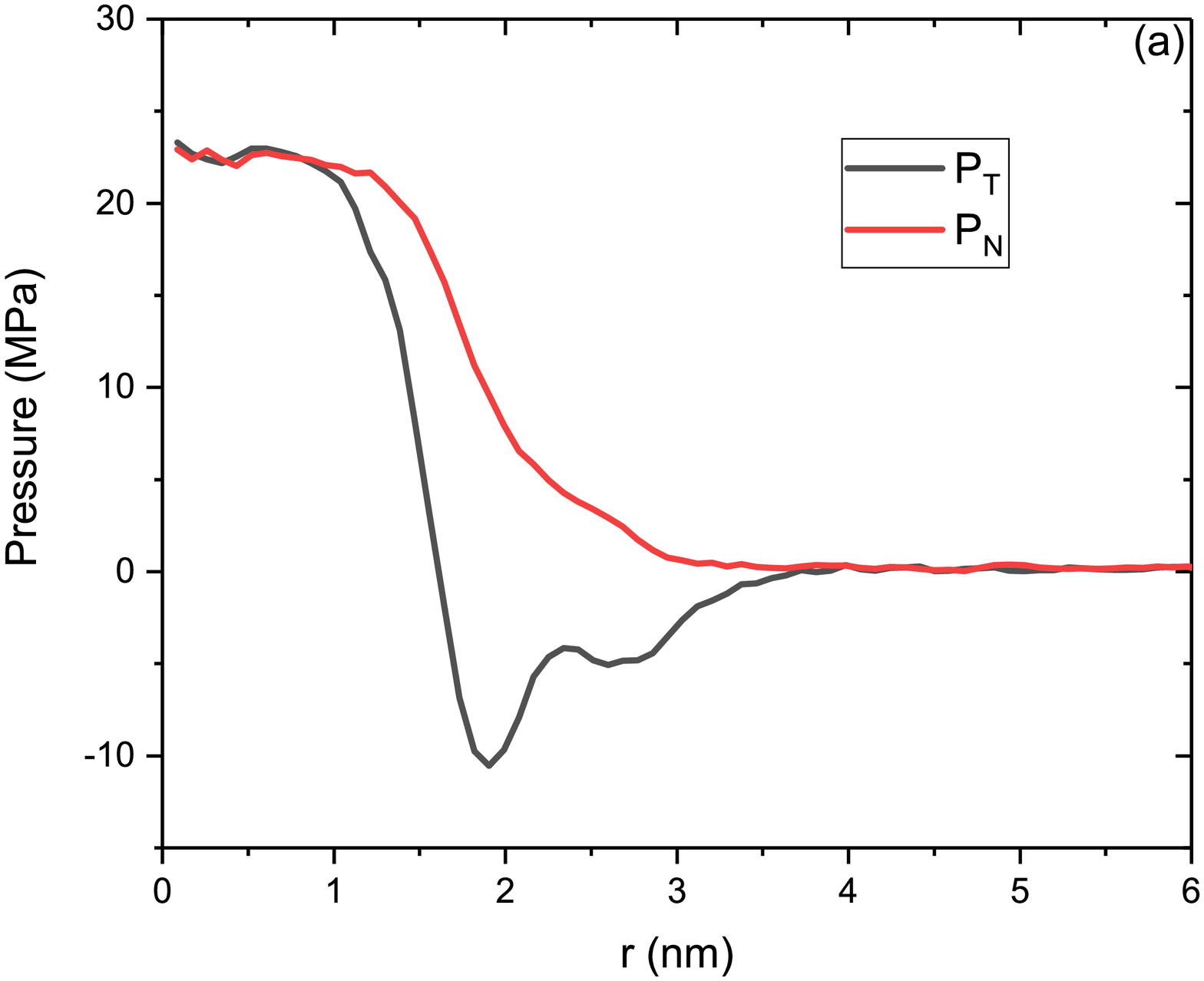}
\includegraphics[scale=0.35]{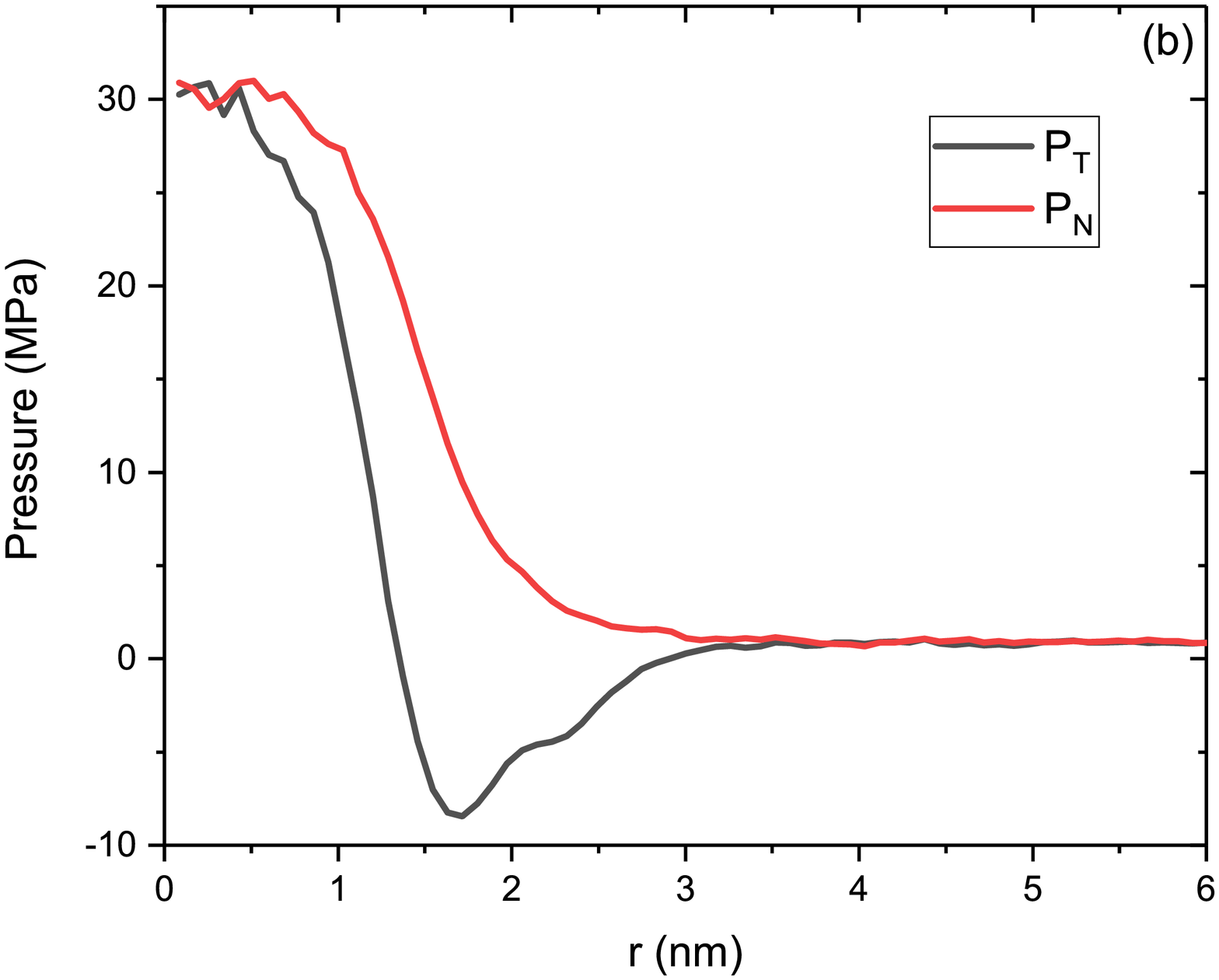}
\includegraphics[scale=0.35]{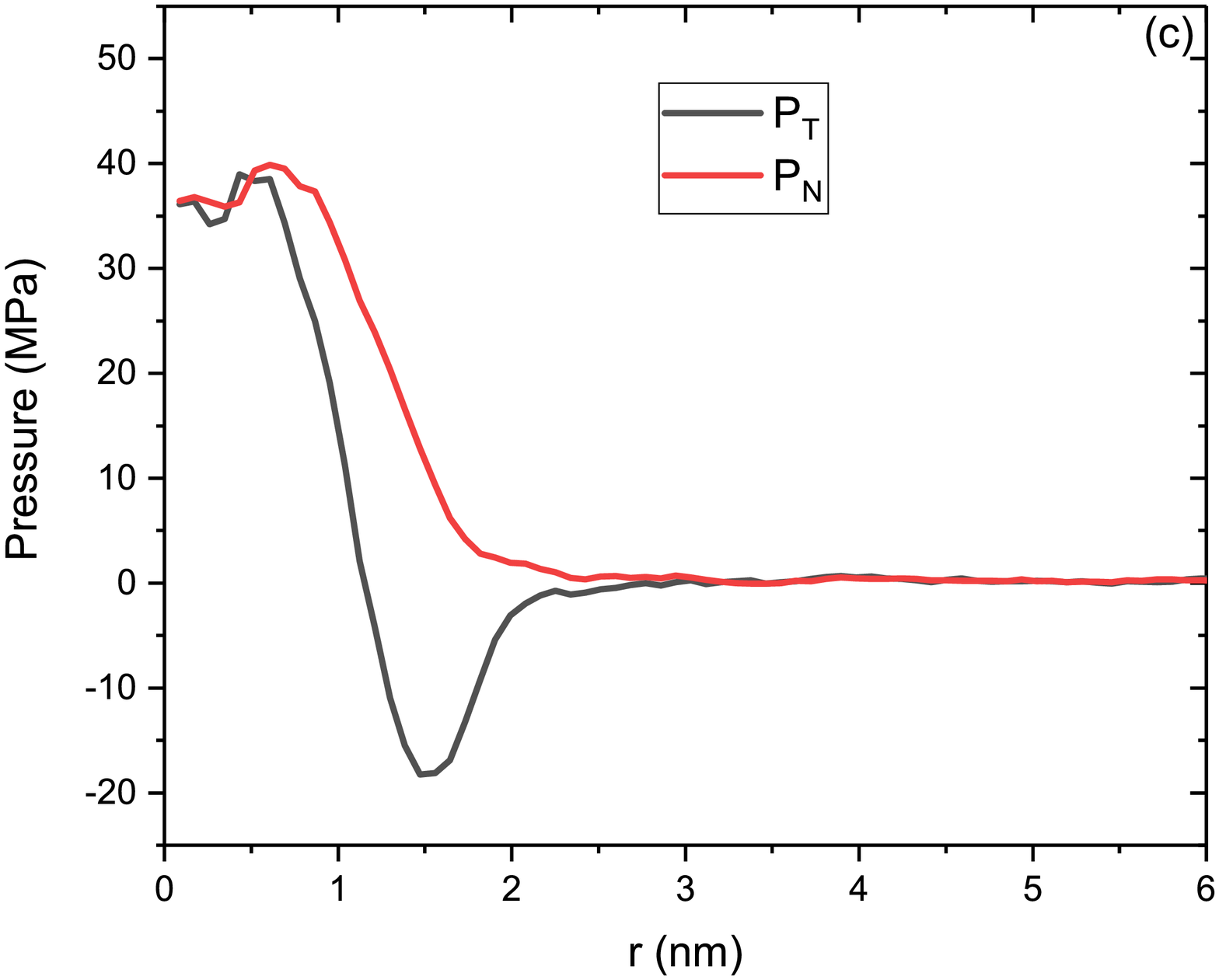}
\caption{Pressure components as a function of the distance from the center of a 2 nm water droplet in \emph{n}-hexane in CG FF simulations, including (a) MARTINI FF, (b) MARTINI FF with polarized
water model and (c) SAFT FF models.}
\label{fig:CG2nmpressure}
\end{figure}

Although it is widely accepted that the Laplace law, relating the pressure jump across a curved interface to its curvature, fails for
nanodroplets, the limit of the Laplace law validity is controversial. Takahashi
and Morita concluded that this limit is less than 1 nm.\cite{RN8} For liquid droplets in vapor environment, thid limit was found to be between 5--10 nm.\cite{RN10,RN12}  Figures \ref{fig:AA2nmpressure} and \ref{fig:CG2nmpressure} show the normal and tangential components of the pressure tensor for a 2 nm water droplet in \emph{n}-hexane.  
%We see that the local normal component and tangential component of the pressure tensor are equal in water and \emph{n}-hexane bulk phases, which are 
%the same in both atomistic and CG simulations, although the fluctuation is large for small nanodroplet. 
We see negative
peaks in the tangent pressure profile at the interface in all simulations, indicating that the interface
is under compression. Similar to the planar interface in atomistic
simulations, we find a small peak on the water side of the tangent pressure in the droplet atomistic 
simulations.
 The pressure in the water
droplet is greater than that in the hexane phase, which is 
consistent with the Laplace law. Comparing Figures \ref{fig:AA2nmpressure} and \ref{fig:CG2nmpressure},
we find that the inner pressure in the atomistic simulations is higher
than that in the CG simulations. In addition, electrostatic interactions in the MARTINI FF
slightly increase the inner pressure, as shown
in Figure \ref{fig:CG2nmpressure}(b). 
%The inner part of the pressure tensor is not so smooth
%in CG simulation (see Figure \ref{fig:CG2nmpressure}(a)), which is because the local density
%fluctuation is strong in small nanodroplet. This is improved with the polarized MARTINI CG water model.

Table \ref{tab:2nmif} lists the interfacial tensions of a 2 nm water droplet in \emph{n}-hexane obtained from the atomistic and CG simulations. Both atomistic models result in a similar interfacial tension, which is smaller than the interfacial tension of the planar interface. 
%Interestingly, these calculated interfacial tensions of curved interface are also very close,
%although the atomistic FFs are different. 
Similar to the planar interface, the interfacial tension calculated with the MARTINI CG FFs is much smaller than that provided by the corresponding atomistic simulation.
The SAFT CG FF, which is able to reproduce the interfacial tension of the
planar interface, also results in a nearly 50 \% smaller interfacial tension than that in the atomistic simulations.  
%Perhaps it is because the solubility of water in \emph{n}-hexane in SAFT CG FF is overestimated. And this effect is enhanced at nanoscale.

\begin{table}
\caption{The interfacial tensions $\gamma_2$ (R=2 nm) of a water droplet in \emph{n}-hexane for
various atomistic and CG FFs at 310 K.}
\begin{tabular}{cc}
\hline
Model & Interfacial tension (mN/m)\\
\hline
atomistic TIP4P2005+TraPPE  & 47.0$\pm$1.1\\
atomistic TIP4P2005+OPLS-AA  & 47.2$\pm$1.9\\
CG MARTINI & 21.4$\pm$2.0\\
CG Polarized MARTINI &
23.9$\pm$2.2\\
CG SAFT & 27.7$\pm$1.1\\
\hline
\label{tab:2nmif}
\end{tabular}
\end{table}

\section{estimation of the interaction
parameters in CG FF using machine learning}\label{ML_methods}

Our results in the previous section show that the MARTINI CG FF cannot reproduce the
interfacial tension and density profile near the interface observed in our atomistic simulations. 
The SAFT CG FF can predict the interfacial tension of the planar interface but underestimates the interfacial tension of the curved interface by almost 50\%. In addition, we find
that the SAFT CG FF overestimates the solubility of water in \emph{n}-hexane. 
We hypothesize that one reason for the poor performance of the above tested CG FFs is  that they use a relatively high degree of coarse graining. We therefore propose using the SDK CG FF \cite{RN23} as it allows a lower degree of coarse-graining. The current SDK model does not define the parameters of the water-hexane potential for the low coarse-graining degree water model. 
We note that there is an SDK FF for the high coarse-graining degree water model, but we find that this water model leads to crystallization of large water droplets.  

In the remainder of this paper, we propose a new approach for learning coarse-grained potentials, apply it to estimating parameters in the water-hexane potential under the SDK CG FF framework, and test the resulting model for the water-hexane system against atomistic simulations. In this work, we use the 1:2 water model (one CG water bead represents two water molecules) and the 1:3 hexane model. The  
potential between CG water and hexane beads is given as
\begin{equation}
U_{\text{w-h}}= \left( \frac{\lambda_{r}}{\lambda_{r} - \lambda_{a}} \right)
\left( \frac{\lambda_{r}}{\lambda_{a}} \right)^\frac{\lambda_{a}}{\lambda_{r} - \lambda_{a}}\varepsilon
\left( \left( \frac{\sigma}{r} \right)^{\lambda_{r}}-\left( \frac{\sigma}{r} \right)^{\lambda_{a}} \right),
\label{eq:potential}
\end{equation}
where \(\lambda_{r}\) and \(\lambda_{a}\) are repulsive and attractive
exponents, $\varepsilon$ is the energy parameter, and $\sigma$ is the core diameter.  
The potentials $U_{\text{w-w}}$ and $U_{\text{h-h}}$ between water-water and hexane-hexane beads have the same form, with parameters  $\lambda_r$, $\lambda_a$, $\sigma$, and $\varepsilon$ listed in Table \ref{tab:cgpar}.
In the original SDK framework, there are only two combinations of $\lambda_r$ and $\lambda_a$, (12,4) and (9,6). The former combination results in a sharper interface because of the larger repulsive force corresponding to $\lambda_r=12$. In the atomistic simulations, we observe a relatively sharp water-hexane interface. Therefore, in the $U_{\text{w-h}}$  potential, we set  $\lambda_r=12$ and $\lambda_a = 4$. 
Next, we learn the $\sigma$ and $\varepsilon$ parameters in the $U_{\text{w-h}}$  potential using the surface tension of the planar and curved water-hexane interfaces as target properties. 
% We select $\sigma$ and $\varepsilon$ as unknown parameters because the CG model is more sensitive with respect to these two parameters as compared to  $\lambda_r$ and $\lambda_a$.  

\begin{table}
\caption{The CG interaction parameters of water and hexane.\cite{RN23,RN280}}
\begin{tabular}{ccccc}
\hline
CG model & $\lambda_{r}$ & $\lambda_{a}$ & $\varepsilon$ (kcal/mol) & $\sigma$(nm)\\ 
\hline
Water & 9 &6 & 0.7050 & 0.2908\\
Hexane & 9 & 6 & 0.4690 & 0.4585\\
\hline
\label{tab:cgpar}
\end{tabular}
\end{table}

We define the parameter vector $\theta=(\sigma,\varepsilon)^T$ and use  
polynomial regression (PR) \citep{RN77,RN75,RN76} to construct a surrogate model of the interfacial tension as a function of $\theta$. 
%Specifically, we use the generalized polynomial chaos
%(gPC)\cite{RN77} to construct a surrogate model of the
%interfacial tension as a function of parameters in the potential. 
%The gPC regression\cite{RN77}
%uses a linear combination of a set of special basis functions (Legendre
%polynomial in our case) defined in the parameter space to represent the
%quantity of interest (the interfacial tension in our case). 
PR uses a linear combination of a set of orthogonal basis functions of $\theta$
 to represent the quantity of interest (QoI) $f$: 
\begin{equation}
  \label{eq:gpc}
f(\theta) = \sum_{i=1}^N c_i\psi_i(\theta), 
\end{equation}
where $\psi_i$ are basis functions (Legendre polynomials) and $c_i$ are
constant coefficients.  Here, $f$ is the interfacial tension obtained from the atomistic simulations.
%When samples of $\theta$ in the parameter space 
%can be drawn with equal probability, the basis functions $\psi_i$ are chosen as 
%Legendre polynomials, and the gPC method is equivalent to the conventional 
%polynomial regression (including interaction terms) in statistics. In this work,
%the parameter space is a subset in $\mathbb{R}^2$ in the form of 
%$[a,b]\times [c,d]$, and we assume no prior bias when sampling this parameter 
%space, so $\psi_i$ are Legendre polynomials. 

%determined fusing a surrogate model for the interfacial tension as a function of these two parameters.

 We search parameters in the space $\sigma \in[\sigma_{min}, \sigma_{max}]$ and $\varepsilon \in
[\varepsilon_{min}, \varepsilon_{max}]$  
%To construct the surrogate model, we 
and treat $\sigma$ and $\varepsilon$ as independent uniform
random variables given by

\begin{eqnarray}\label{random_variables}
\left(
\begin{array}{c}
{\sigma}\\
{\varepsilon}
\end{array}\right)
=
\left(
\begin{array}{c}
\overline{\sigma}\\
\overline{\varepsilon}
\end{array}\right)
+\left( \delta_{\sigma},\delta_{\varepsilon} \right)\cdot
\left(
\begin{array}{cc}
{\xi_{1}} & 0 \\
0 & {\xi_{2}} 
\end{array}\right),
\end{eqnarray}
where
\(\left( \overline{\sigma},\overline{\varepsilon} \right) = (0.5,\ 0.225)\) 
are the parameter means and \((\xi_{1},\xi_{2})\)
are independent random variables uniformly distributed on
\(\lbrack - 1,1\rbrack\).
The $\overline{\sigma}$ and $\overline{\varepsilon}$ values are defined as an average of $\sigma$ and $\varepsilon$ in water-water and hexane-hexane potentials, respectively. 
The parameters $\delta_{\sigma}=0.1$ and $\delta_{\varepsilon} = 0.035$ are found as $\delta_\sigma = (\sigma_{max}-\overline{\sigma})$ and  $\delta_\varepsilon = (\varepsilon_{max}-\overline{\varepsilon})$, where $\sigma_{max} = 0.6$ is the maximum size of the water-hexane molecule cluster, and $\varepsilon_{max} = 0.26$  is the interaction energy between water and hexane.  $\sigma_{min}$ and  $\varepsilon_{min}$ are computed as $\sigma_{min} = 2\overline{\sigma}-\sigma_{max}$ and $\varepsilon_{min} = 2\overline{\varepsilon}-\varepsilon_{max}$, respectively.
We generate 49 samples of $\xi_1$ and $\xi_2$ using the sparse grids method~\cite{RN78} with 
one-dimensional Gaussian quadrature points and the tensor product rule (i.e., the
number of samples is equal to $7^d$, where $7$ is the number of one-dimensional
quadrature points and $d$ is the number of unknown parameters). We then
compute $(\sigma, \varepsilon)$ for each sample $(\xi_1, \xi_2)$ from Eq (\ref{random_variables}), simulate the
flat interface using the CG model for these values of $(\sigma, \varepsilon)$, and compute the corresponding interfacial tension.
%in each simulation.  
The values of the interfacial tension are used to estimate the  coefficients $c_i$ in the PR surrogate model (\ref{eq:gpc}) 
based on the probabilistic collocation method.\cite{RN79}
%computed from the realizations
We find that the interfacial tension changes smoothly in the considered parameter space and the relative error of the surrogate model,   based on 10-fold cross validation,\cite{RN303} is less than 1\%.  

\begin{figure}
\includegraphics[scale=0.4]{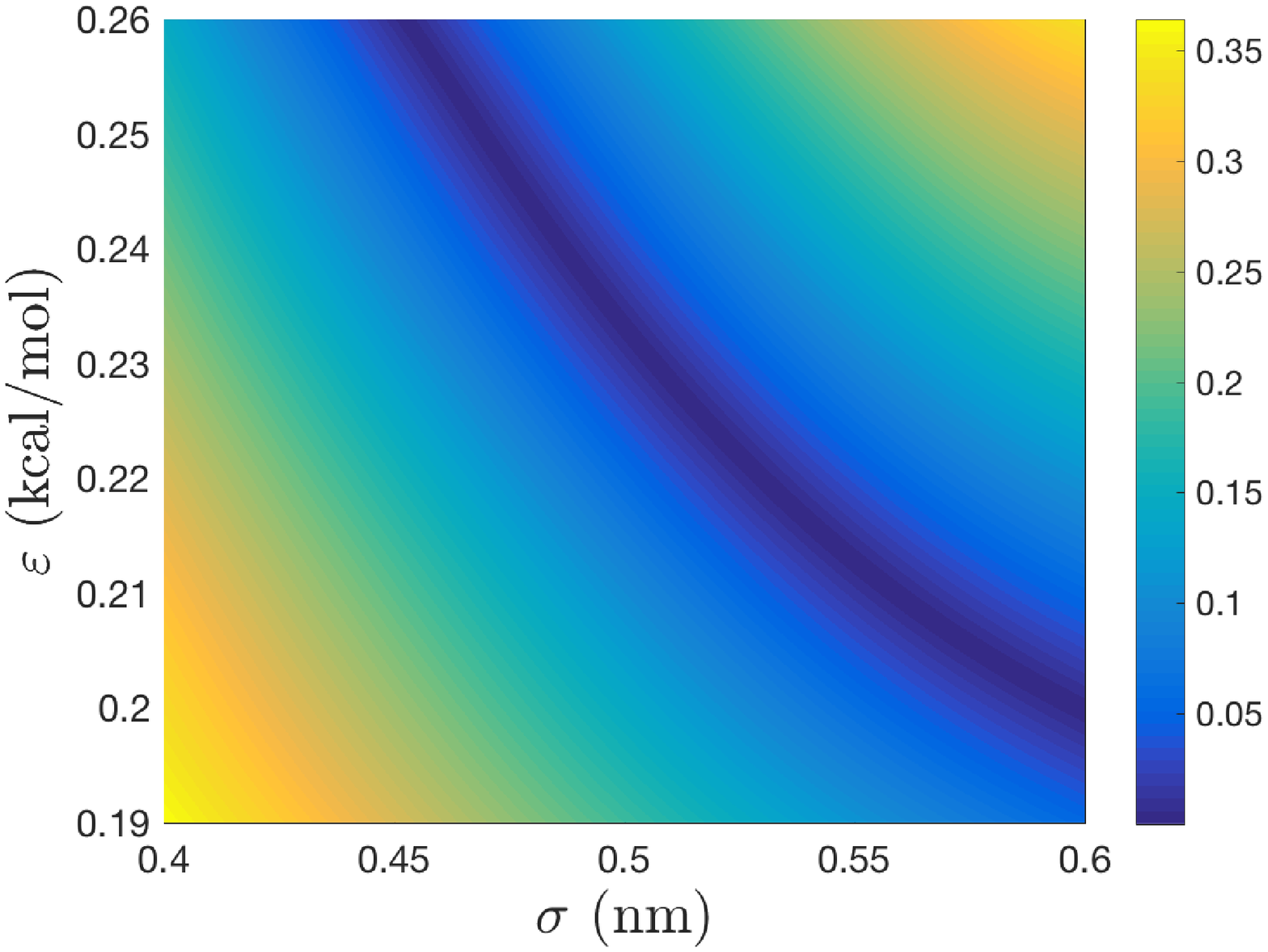}
\includegraphics[scale=0.4]{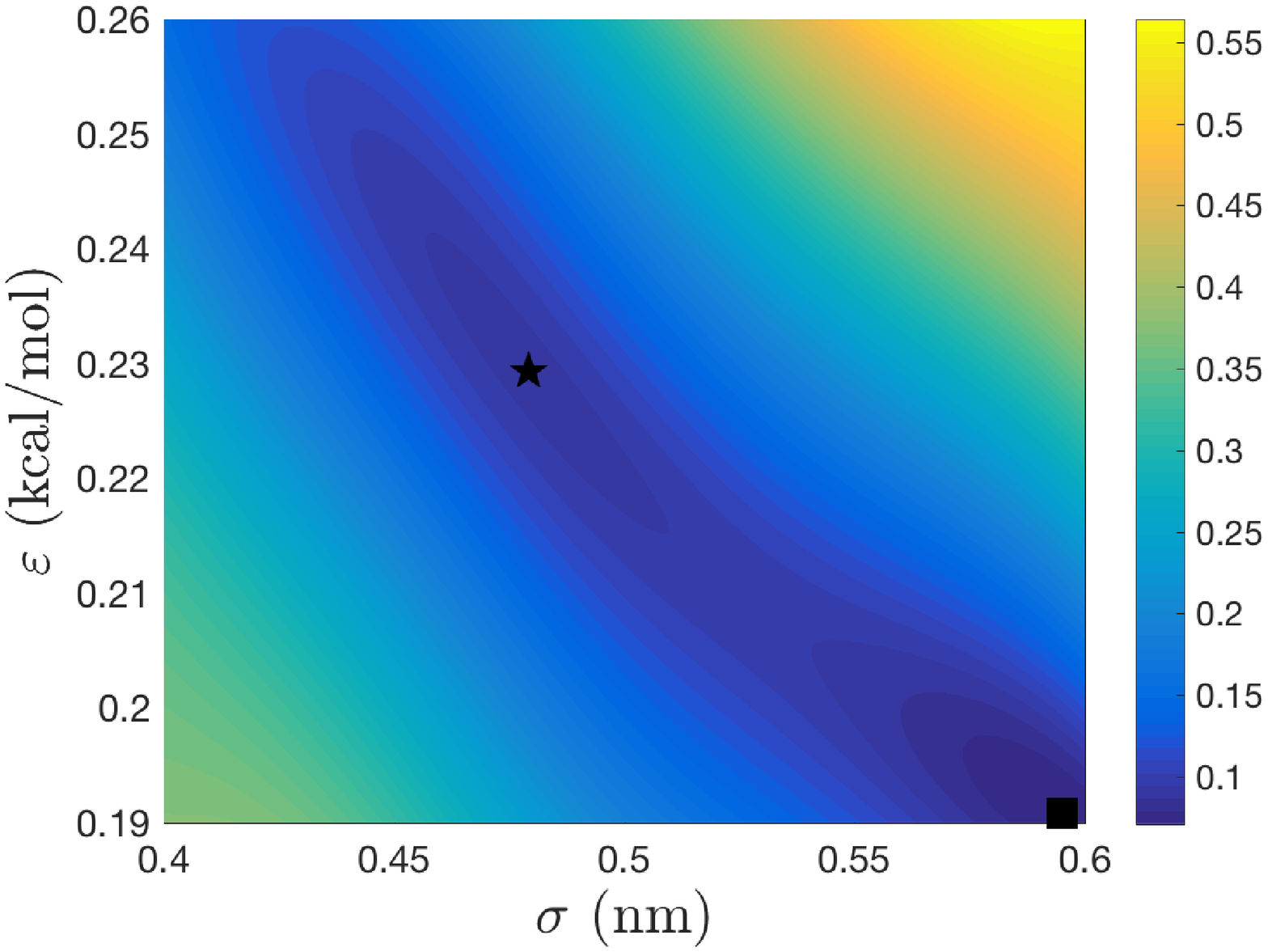}
\caption{
%Response surfaces describing the relative error of the
%interfacial tension in the parameter space for MIE 12-4 potential
%between water and hexane. (a) One target. (b) Two targets.
(a) $\frac{f_0(\sigma,\varepsilon)-\gamma_0}{\gamma_0}$ and (b) $\sqrt {\left( \frac{f_0(\sigma,\varepsilon)-\gamma_0}{\gamma_0} \right)^2 
+ \left( \frac{f_2(\sigma,\varepsilon)-\gamma_2}{\gamma_2}\right)^2 }$ versus $\sigma$ and $\varepsilon$.
}
\label{fig:response}
\end{figure}

Finally, the surrogate model is used to find parameters \(\sigma\) and \(\varepsilon\)  that correspond to the interfacial 
tension of the planar water-hexane interface in the atomistic simulation.
Figure \ref{fig:response}(a) shows $\frac{f_0(\sigma,\varepsilon)-\gamma_0}{\gamma_0}$ ($f_0(\sigma,\varepsilon)$ is the surrogate model for the interfacial tension of the planar interface, and $\gamma_0$ is the interfacial tension of the planar interface obtained from the atomistic simulation) as a function of $\sigma$ and
$\varepsilon$. 
%\textcolor{red}{It is not clear. How can you have a response surface for the error? You only described a response surface for the tension.}
%It is interesting that the optimal parameters lie on one curve. 
There is an infinite number of pairs  \(\left( \sigma,\varepsilon \right)\)
that generate $\gamma_0$ lying on the curve  $\frac{f_0(\sigma,\varepsilon)-\gamma_0}{\gamma_0} = 0$. To make parameterization unique, we select the interfacial tension of a 2 nm
water droplet in hexane ($\gamma(2) = \gamma_2 = 47 mN/m$) as an additional constraint. We use the same 49 samples of random
variables and the corresponding  $\sigma$ and $\varepsilon$ to simulate a water
droplet in hexane with the CG model. To reduce the statistical error caused by the thermal
fluctuations of a small droplet, every sample is averaged over five
independent CG simulations. These simulations are used to construct the surrogate model of the surface tension of a  2 nm droplet $f_2(\sigma,\varepsilon)$. Then, the optimal $\sigma$ and $\varepsilon$ are determined
by solving the minimization problem:
\begin{equation}
(\sigma,\varepsilon) = \min_{\sigma,\varepsilon} \left[ \sqrt {\left( \frac{f_0(\sigma,\varepsilon)-\gamma_0}{\gamma_0} \right)^2 
+ \left( \frac{f_2(\sigma,\varepsilon)-\gamma_2}{\gamma_2}\right)^2 } \right].
\end{equation}
Figure \ref{fig:response}(b) shows $\sqrt {\left( \frac{f_0(\sigma,\varepsilon)-\gamma_0}{\gamma_0} \right)^2 
+ \left( \frac{f_2(\sigma,\varepsilon)-\gamma_2}{\gamma_2}\right)^2 }$ as a function of $\sigma$ and $\varepsilon$. It can be seen that there are two sets of optimal parameters:
\(\varepsilon = 0.23\) kcal/mol and \(\sigma = 0.48\) nm (the star); and
\(\varepsilon = 0.19\) kcal/mol and \(\sigma = 0.59\) nm (the square). The
difference between the response surface values at these two points is
less than 2\%, which is within the range of fluctuations observed in the CG  
simulations.

\begin{figure}
\includegraphics[scale=0.4]{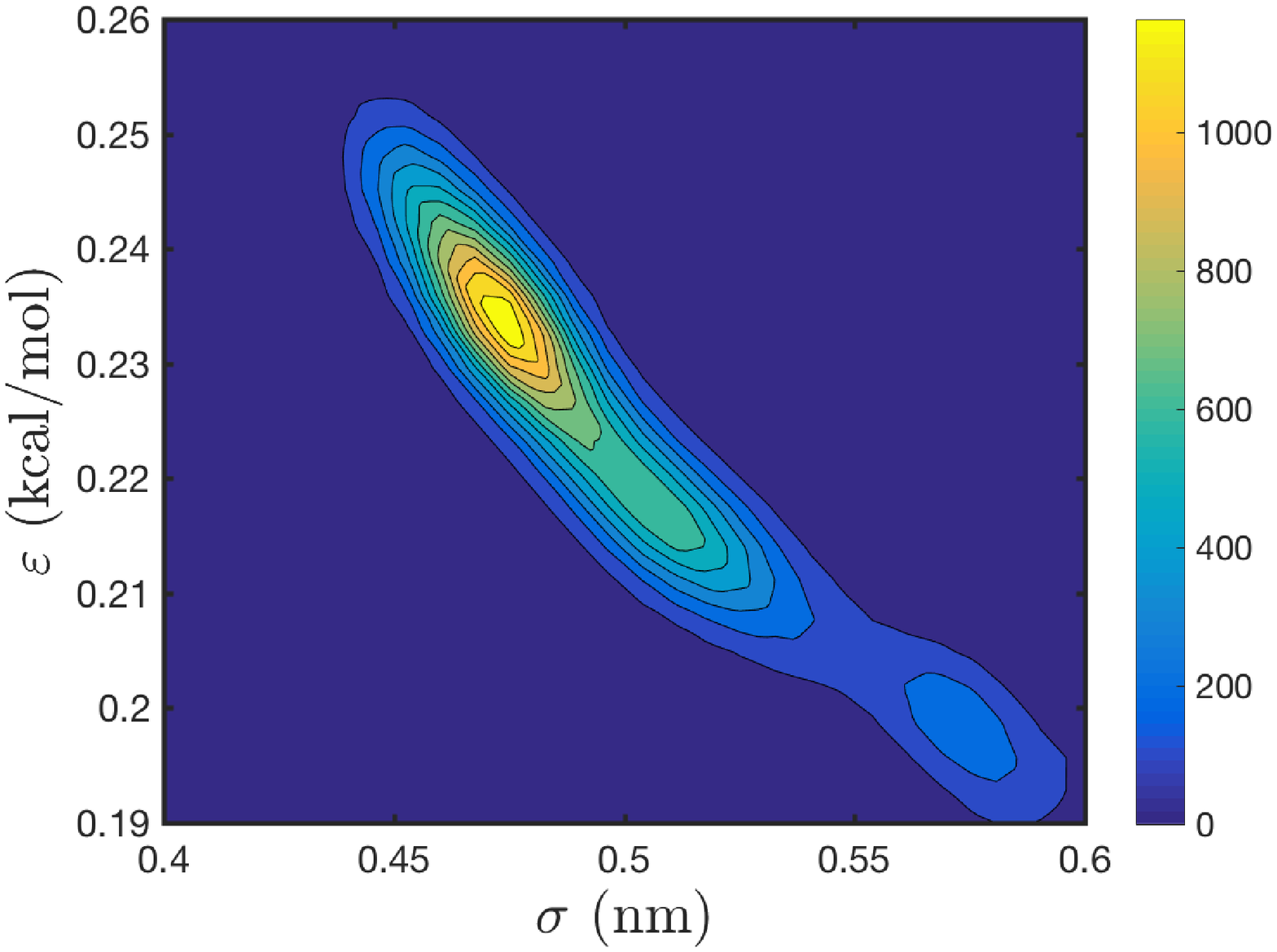}
\includegraphics[scale=0.4]{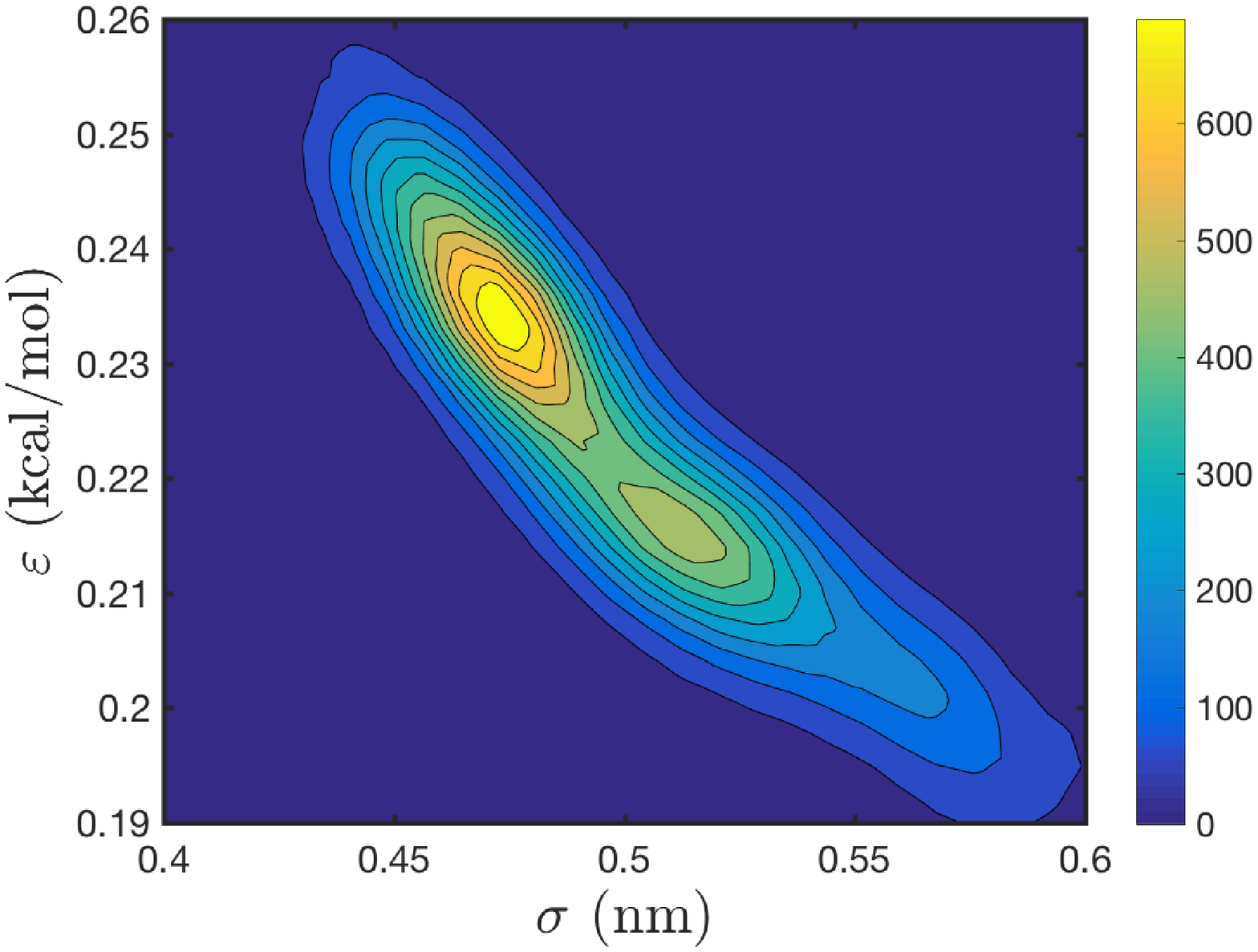}
\caption{PDF function of the optimal parameter via UQ analysis of the
CG interaction with various input noises. (a) 10\% Gaussian noise and (b)
20\% Gaussian noise.}
\label{fig:pdfs}
\end{figure}

Fluctuations in the value of the interfacial tension computed from the CG simulations are due to thermal fluctuations. When the
interfacial tension is used as a target to estimate 
parameters, these fluctuations (which can be treated as uncertainty) should be transferred to parameters. 
This requires knowledge of the interfacial tension sensitivity with respect to the parameters $\sigma$ and $\varepsilon$. 
%Therefore, we perform UQ analysis of the CG interaction parameter space. 
%The same 49 samples of input parameters \((\varepsilon,\ \sigma)\) with additional Gaussian noise in the output of \textcolor{red}{CG (???)} MD simulations (planar and curved interface) are tested. 
To perform the sensitivity analysis, we add 10\% and 20\% Gaussian noise to the values of the interfacial tension obtained from the 49 CG 
simulations, construct the surrogate model, and determine the optimal parameter set \((\varepsilon, \sigma)\) as described above.
 We repeat this procedure 100,000 times and compute the probability density function (PDF) of the optimal \((\varepsilon,\ \sigma)\). 
Figure \ref{fig:pdfs} shows the PDFs of the \((\varepsilon, \sigma)\) parameter set. We 
see that the PDF maximum is
at  the position \(\left( \varepsilon,\sigma \right) = (0.23,0.48)\)
for both 10\% and 20\% added noise, with the peak in the 10\% case being steeper than that in the 20\%
case. This is because the smaller noise (uncertainty) in the surface tension leads to a more certain estimate  of the
optimal parameters. The important feature of Figure \ref{fig:pdfs} is that, unlike Figure \ref{fig:response},
it predicts a unique set of the optimal parameters. The square in Figure \ref{fig:response} corresponds to a
point with very small probability in Figure \ref{fig:pdfs}. 
With  \(\left( \varepsilon,\sigma \right) = (0.23,0.48)\), the CG model produces an interfacial tension of 53.2 mN/m for the planar interface and 42.0 mN/m for the curved interface: these are within 11\% of the values obtained in atomistic simulations. Finally, we test the CG model with  \(\left( \varepsilon,\sigma \right) = (0.23,0.48)\), by simulating a 3 nm water droplet in \emph{n}-hexane and find that the interfacial tension is 45.1 mN/m, which is within 9\% of the 49.1 mN/m interfacial tension value computed from the atomistic simulation of the 3 nm water droplet. 

%Here, we use sparse grids method to construct a polynomial regression surrogate model.
%Other efficient approaches can be used in this step, e.g.,
%compressive-sensing-based method, \cite{RN80,RN81,RN82} especially when
%the number of parameters is large. In addition, the PDF of the optimal
%parameters (shown in Figure \ref{fig:pdfs}) can be computed more efficiently using
%Bayesian inference.\cite{RN75, RN83}
\begin{figure}
\includegraphics[scale=0.35]{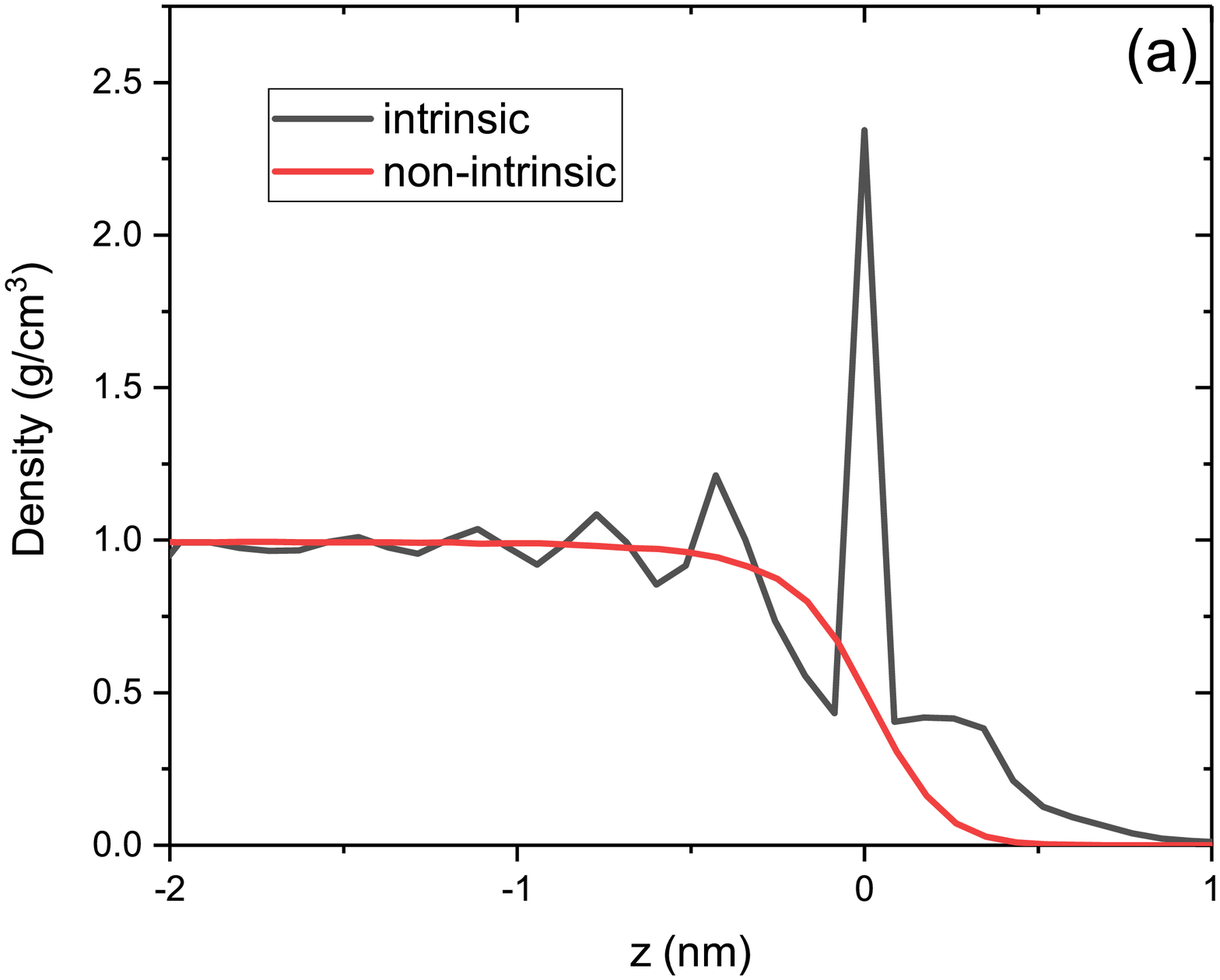}
\includegraphics[scale=0.35]{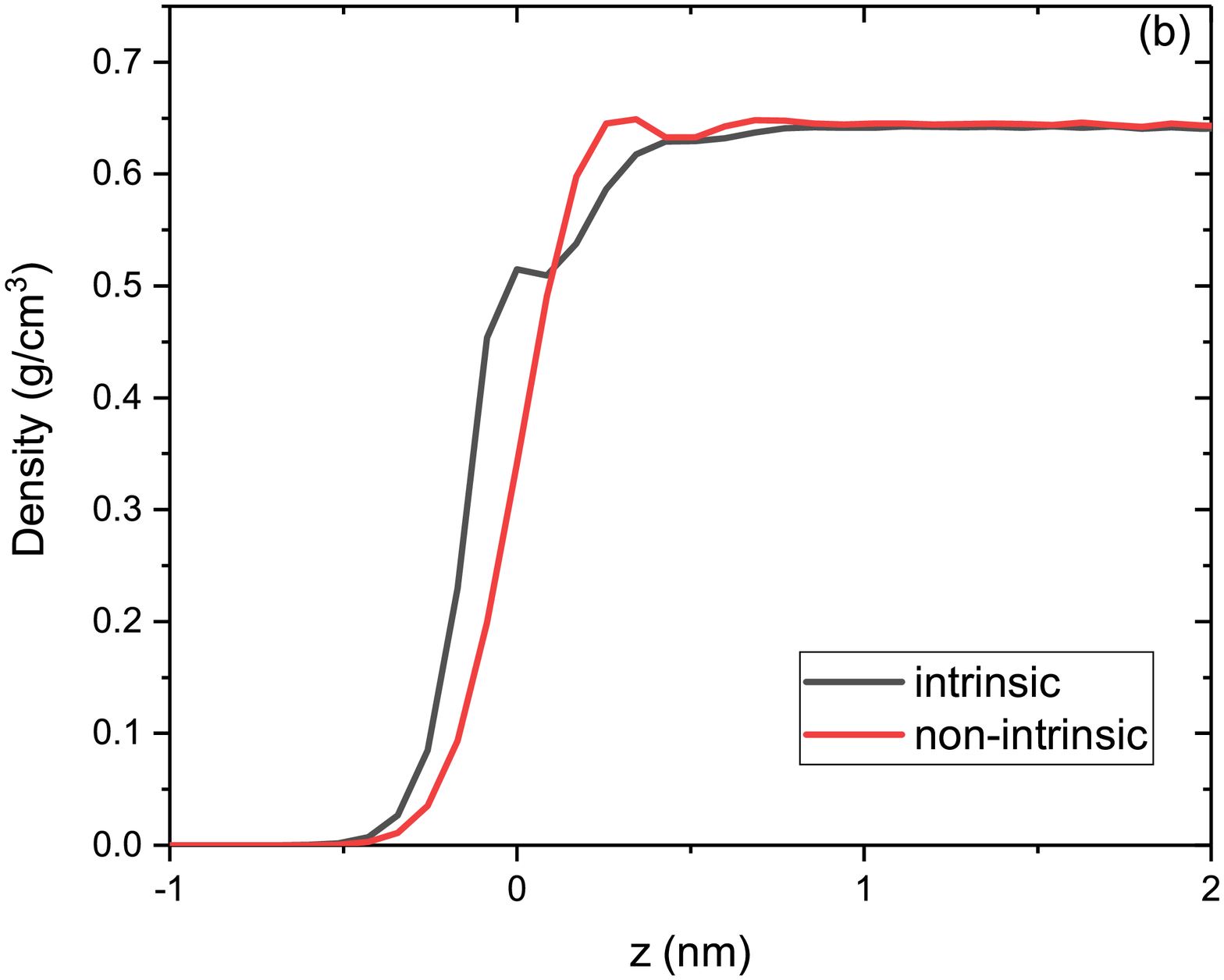}
\includegraphics[scale=0.35]{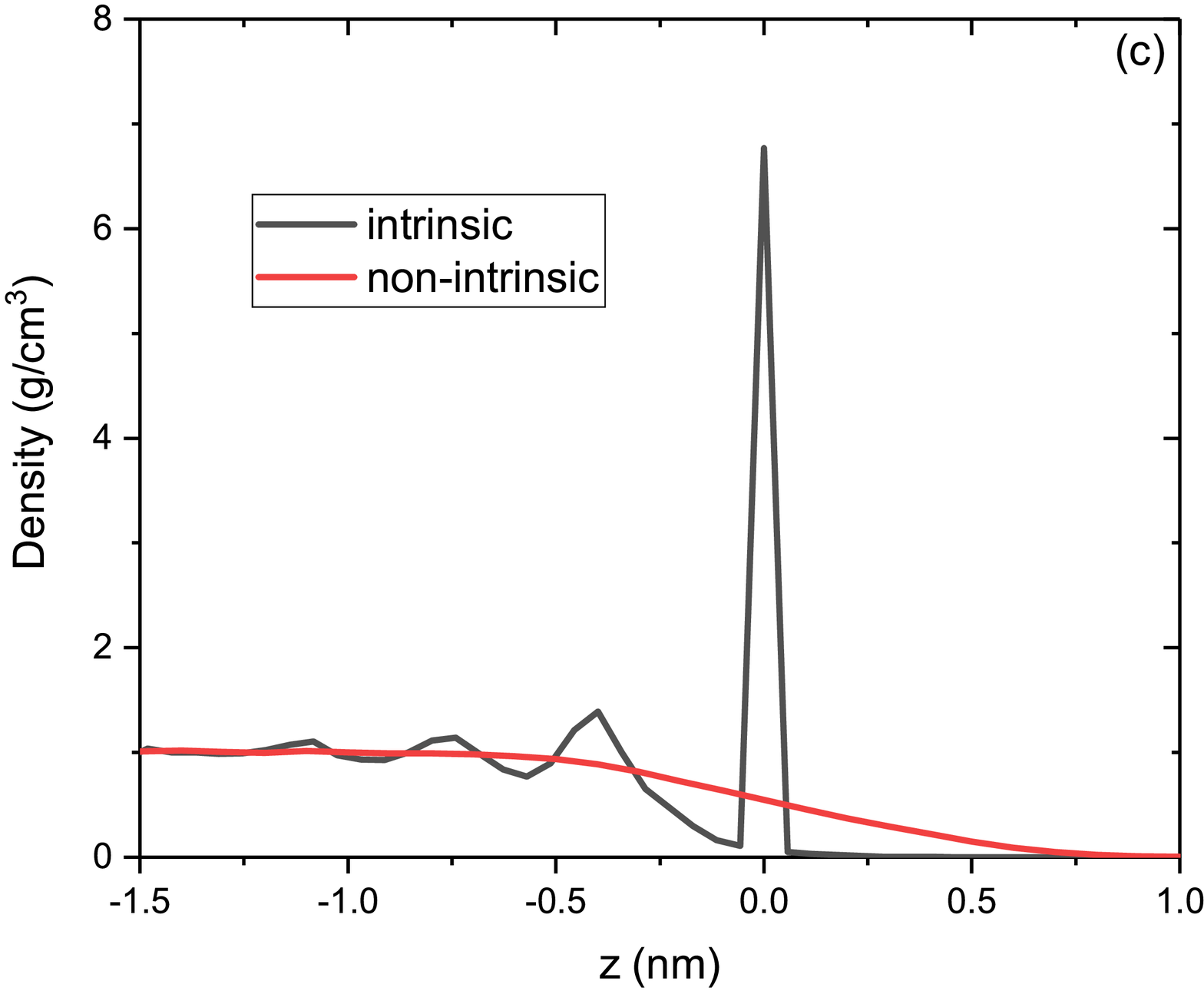}
\caption{The intrinsic and non-intrinsic 
density profiles of a water-hexane interface obtained with the CG FF with learned $\sigma$ and $\varepsilon$. (a) Water at
water-hexane planar interface. (b) Hexane at the water-hexane planar
interface. (c) Water at the water-hexane curved interface.}
\label{fig:mlmodel}
\end{figure}

The intrinsic and non-intrinsic density profiles for planar and curved interfaces obtained from the CG model with  \(\left( \varepsilon,\sigma \right) = (0.23,0.48)\) are presented in Figure \ref{fig:mlmodel}. The comparison of Figures \ref{fig:mlmodel}(a) and (b) with Figures \ref{fig:AA_density}(a) and (b) shows that the width of the planar interface is very similar in the atomistic and CG simulations, as well as in the resulting non-intrinsic density profiles for both water and hexane.
For the planar interface, there are three peaks
in the ``CG'' intrinsic density profile of water, while only two peaks are observed in the ``atomistic'' intrinsic water density profile. The CG FF produces a longer-range ordered structure because it uses a larger cutoff than the atomistic FF. On the other hand, the locations and magnitudes of the first two peaks in the CG density profile are close to those in the atomistic simulations. 
The comparison of Figures \ref{fig:mlmodel}(c) and  \ref{figAA2nmdensity} also demonstrates good agreement between the 
the intrinsic and non-intrinsic density profiles of a 2 nm water droplet (i.e., curved interface) obtained with our CG model and the atomistic models in terms of the bulk water density, interface width, and structure.
There are some disagreements in the intrinsic density profiles of hexane in the CG and atomistic simulations. There is a relatively small peak in the intrinsic atomistic hexane density profile and no apparent
 peak in the CG intrinsic hexane density profile. 
 This disagreement is caused by the coarse-graining of the one-site CG water and two-site CG hexane models. 
 %This is consistent with the above conclusion that the geometry of CG model would affect the local structure of the interface in the simulation. 

%\textcolor{red}{We note that the SAFT CG FF overestimates the width of the water-hexane interface and the density of water in hexane, see Figure \ref{figCG2nmdensity}. } \textcolor{blue}{I do not see this from Figure 7.} \textcolor{red}{It can be concluded when we compare the non-intrinsic density profile in Figure 7 to atomic results in Figure 2, but currently it seems that we do not need it any more. I move it to page 6. }
 
%SAFT CG FF, which means that the two CG systems have similar interface structure. However, compared
%with the non-intrinsic density profile in SAFT CG FF, it is seen that
%the width of interface predicted by our parameter sets is much narrow,
%which is close to the atomistic simulation. As mentioned above, the
%density of water in n-hexane predicted by SAFT bio 2 CG model is
%overestimated. According to experiment result, the density of water in
%n-hexane is less than 0.058 kg/m\textsuperscript{3} at 298
%K.\cite{RN84}

\section{Conclusion}\label{conclusions}
We proposed a new approach for parameterization of CG FF and used it to parameterize the SDK CG FF for water-hexane systems with the interfacial tension of planar and curved interfaces as target properties. We demonstrated that the proposed model significantly improves the accuracy  of CG predictions of the coexisting densities, interface structure, and interfacial tension for planar and
curved water-hexane interfaces. 

We started this work by testing two atomistic FFs and three CG FFs for water-hexane systems, including a planar interface separating water and hexane and a water droplet in hexane. We found that the results of both atomistic FFs agree well with experiments and later used them as references.  Next, we studied three popular CG FFs and found that none of them can accurately reproduce the interfacial structure and interfacial tension of planar and/or curved water-hexane interfaces. 

Finally, we proposed a new approach for learning parameters in a CG FF with respect of target properties and used it to parameterize the SDK CG FF for water-hexane system using the surface tension of a planar interface and a 2 nm water droplet in hexane as target properties. 
We found that the proposed SDK CG FF can accurately describe the interface structure and predict the curvature-dependent interfacial tension of the water-hexane system. For the latter, we demonstrated that the CG model can estimate the interfacial tension of a 3 nm water droplet (in addition to the surface tension of a 2 nm droplet and planar interface that we used to train the CG model). 
The proposed approach can be easily generalized for learning parameters in other CG models of complex fluids using appropriate target properties.

%As a result, the machine learning CG potentials open the way to offering more quantitative
%predictions of interfacial systems. Our results are helpful for
%designing and application of CG FFs for interface systems and the
%treatment can be extended to understand the complex physical and
%biological interface as nanoemulsions, micelles and vesicles formed in
%solutions. 

\section*{\label{sec:level1}Acknowledgements}
This work was supported by the U.S. Department of Energy, Office of
Science, Office of Advanced Scientific Computing Research. Pacific Northwest National Laboratory is operated by Battelle for the DOE under Contract DE-AC05-76RL01830.

\appendix
\section{Non-intrinsic and intrinsic density profiles calculations}\label{App_densities}

Non-intrinsic density profiles are calculated along the direction normal to the interface
using a bin size of 0.2 nm. The non-intrinsic density is averaged within each bin over time. However, these non-intrinsic density profiles are subject to capillary waves due to thermal fluctuations. Recently, the intrinsic sampling method was developed to obtain the intrinsic density of liquid-vapor and liquid-liquid interfaces.\cite{RN54,RN55,RN56,RN57} The
capillary wave theory provides the dependence of a density profile on the
interfacial area. This theory postulates that a liquid surface can be
represented by the intrinsic surface
\(z = \ \xi(\mathbf{R},q)\). The intrinsic surface is defined in terms
of a surface layer of $N$ molecules or molecular
sites with coordinates \(\mathbf{R} = \left( x,y \right)\) on the
transverse plane. The vector $\mathbf{R}$ represents the instantaneous molecular border of the coexisting phases and depends on the  the wave vector
cutoff $q$, which defines the surface resolution. For a given 
$q$, the intrinsic density profile is defined as
\begin{equation}
\rho\left( z \right) = \frac{1}{A}\left\langle \sum_{i}^{}{\delta(z - z_{i} + \gamma (x_{i},y_{i}))} \right\rangle,
\end{equation}
where $z$ is the local non-intrinsic interface location, $A$ is the
cross-section area of the interface, 
\(\left\langle \cdots \right\rangle\) is the ensemble statistical
average, and $\gamma$ is the amplitude of the thermal
fluctuations.

We use the ITIM algorithm\cite{RN57,RN58,RN59,RN709} for identifying
the interfacial molecules that are exposed to the opposite phase using a
probe sphere radius of 0.2 nm. The probe sphere
is moved along test lines perpendicular to the plane of the fluid-fluid
interface. Atoms that first encounter the probing ball are
identified as the interfacial atoms, and the corresponding molecules are
identified as the interfacial molecules. This process is repeated over
the entire interfacial area in the simulation. 

\section{Pressure profile and interfacial tension calculations}\label{App_pressure}

%There are three common approaches for computing the interfacial tension at fluid interfaces, including mechanical, thermodynamic, and statistical mechanics approaches.\cite{RN12} 
%In MD simulations,
%the interfacial tension is usually obtained by the statistical
%mechanics approach. Actually, the routes based on statistical
%mechanical approach also borrowed some concepts in the classical
%mechanical approach and thermodynamic approach at continuum level. For example, in the virial route,\cite{RN60} the surface tension %is
%found from the isochoric--isothermal change in free energy due to an
%increase in the interfacial surface area. While the standard
%stress-strain formulae follow from the first-order change in free
%energy due to a deformation of the area and lead to the mechanical
%expression of Kirkwood and Buff.\cite{RN61} In a particle-based
%simulation Ghouri and Makfreyt have indicated the surface tension is
%invariant to the choice of the thermodynamic and the mechanical routes
%for the pressure tensor calculation in spherical
%geometries.\cite{RN10} Ndao \emph{et al.} also compared the
%interfacial tensions for several immiscible binary systems obtained by
%mechanical routes\cite{RN62,RN63} and thermodynamic
%route\cite{RN64} in MD simulation but did not find obvious
%difference between them.\cite{RN65} So in this paper the
%mechanical route is implemented to the calculation of interfacial
%tension. 
The interfacial tension of a planar interface is computed as\cite{RN66}
\begin{equation}
\gamma(z) = \int_{- z/2}^{z/2}{(P_{N}\left( z \right) - P_{T}\left( z \right))dz}.
\label{flat}
\end{equation}

For a spheric droplet, the expression for the surface tension takes the form
\begin{equation}
\gamma(r) = \int_{0}^{\infty}{(P_{N}\left( r \right) - P_{T}\left( r \right))dr},
\label{curved}
\end{equation}
where \(P_{N}\) and \(P_{T}\) are the normal and tangential components
of the pressure tensor along the normal  direction to the surface. We use the Irving-Kirkwood
\cite{RN62} and Vanegas and Ollila\cite{RN67,RN245} approaches for computing pressure components in Eqs (\ref{flat}) and (\ref{curved}), respectively.     
These approaches were originally proposed for pair-wise interactions.  To calculate pressures due to three-body angle potentials, these potentials are  decomposed into pair-wise potentials by the central force decomposition (CFD) method.\cite{RN69}
%The electrostatic interaction in MD simulation is often solved with
%Ewald summation method, which includes reciprocal space electrostatic interaction as many-body interaction. 
%In our calculation of local pressure tensor, 
The many-body electrostatic interactions are
approximated as pairwise interaction. In our pressure calculations, we use a pairwise potential with the 2.0 nm cutoff.

\bibliography{interface}

\begin{thebibliography}{54}%
\makeatletter
\providecommand \@ifxundefined [1]{%
 \@ifx{#1\undefined}
}%
\providecommand \@ifnum [1]{%
 \ifnum #1\expandafter \@firstoftwo
 \else \expandafter \@secondoftwo
 \fi
}%
\providecommand \@ifx [1]{%
 \ifx #1\expandafter \@firstoftwo
 \else \expandafter \@secondoftwo
 \fi
}%
\providecommand \natexlab [1]{#1}%
\providecommand \enquote  [1]{``#1''}%
\providecommand \bibnamefont  [1]{#1}%
\providecommand \bibfnamefont [1]{#1}%
\providecommand \citenamefont [1]{#1}%
\providecommand \href@noop [0]{\@secondoftwo}%
\providecommand \href [0]{\begingroup \@sanitize@url \@href}%
\providecommand \@href[1]{\@@startlink{#1}\@@href}%
\providecommand \@@href[1]{\endgroup#1\@@endlink}%
\providecommand \@sanitize@url [0]{\catcode `\\12\catcode `\$12\catcode
  `\&12\catcode `\#12\catcode `\^12\catcode `\_12\catcode `\%12\relax}%
\providecommand \@@startlink[1]{}%
\providecommand \@@endlink[0]{}%
\providecommand \url  [0]{\begingroup\@sanitize@url \@url }%
\providecommand \@url [1]{\endgroup\@href {#1}{\urlprefix }}%
\providecommand \urlprefix  [0]{URL }%
\providecommand \Eprint [0]{\href }%
\providecommand \doibase [0]{http://dx.doi.org/}%
\providecommand \selectlanguage [0]{\@gobble}%
\providecommand \bibinfo  [0]{\@secondoftwo}%
\providecommand \bibfield  [0]{\@secondoftwo}%
\providecommand \translation [1]{[#1]}%
\providecommand \BibitemOpen [0]{}%
\providecommand \bibitemStop [0]{}%
\providecommand \bibitemNoStop [0]{.\EOS\space}%
\providecommand \EOS [0]{\spacefactor3000\relax}%
\providecommand \BibitemShut  [1]{\csname bibitem#1\endcsname}%
\let\auto@bib@innerbib\@empty
%</preamble>
\bibitem [{\citenamefont {Ishiyama}, \citenamefont {Imamura},\ and\
  \citenamefont {Morita}(2014)}]{RN1}%
  \BibitemOpen
  \bibfield  {author} {\bibinfo {author} {\bibfnamefont {T.}~\bibnamefont
  {Ishiyama}}, \bibinfo {author} {\bibfnamefont {T.}~\bibnamefont {Imamura}}, \
  and\ \bibinfo {author} {\bibfnamefont {A.}~\bibnamefont {Morita}},\ }\href
  {\doibase 10.1021/cr4004133} {\bibfield  {journal} {\bibinfo  {journal}
  {Chem. Rev.}\ }\textbf {\bibinfo {volume} {114}},\ \bibinfo {pages} {8447}
  (\bibinfo {year} {2014})}\BibitemShut {NoStop}%
\bibitem [{\citenamefont {Piradashvili}\ \emph {et~al.}(2016)\citenamefont
  {Piradashvili}, \citenamefont {Alexandrino}, \citenamefont {Wurm},\ and\
  \citenamefont {Landfester}}]{RN2}%
  \BibitemOpen
  \bibfield  {author} {\bibinfo {author} {\bibfnamefont {K.}~\bibnamefont
  {Piradashvili}}, \bibinfo {author} {\bibfnamefont {E.~M.}\ \bibnamefont
  {Alexandrino}}, \bibinfo {author} {\bibfnamefont {F.~R.}\ \bibnamefont
  {Wurm}}, \ and\ \bibinfo {author} {\bibfnamefont {K.}~\bibnamefont
  {Landfester}},\ }\href {\doibase 10.1021/acs.chemrev.5b00567} {\bibfield
  {journal} {\bibinfo  {journal} {Chem. Rev.}\ }\textbf {\bibinfo {volume}
  {116}},\ \bibinfo {pages} {2141} (\bibinfo {year} {2016})}\BibitemShut
  {NoStop}%
\bibitem [{\citenamefont {Liu}, \citenamefont {Li},\ and\ \citenamefont
  {Shao}(2011)}]{RN3}%
  \BibitemOpen
  \bibfield  {author} {\bibinfo {author} {\bibfnamefont {S.~J.}\ \bibnamefont
  {Liu}}, \bibinfo {author} {\bibfnamefont {Q.}~\bibnamefont {Li}}, \ and\
  \bibinfo {author} {\bibfnamefont {Y.~H.}\ \bibnamefont {Shao}},\ }\href
  {\doibase 10.1039/c0cs00168f} {\bibfield  {journal} {\bibinfo  {journal}
  {Chem. Soc. Rev.}\ }\textbf {\bibinfo {volume} {40}},\ \bibinfo {pages}
  {2236} (\bibinfo {year} {2011})}\BibitemShut {NoStop}%
\bibitem [{\citenamefont {van~der Spoel}, \citenamefont {van Maaren},\ and\
  \citenamefont {Caleman}(2012)}]{RN30}%
  \BibitemOpen
  \bibfield  {author} {\bibinfo {author} {\bibfnamefont {D.}~\bibnamefont
  {van~der Spoel}}, \bibinfo {author} {\bibfnamefont {P.~J.}\ \bibnamefont {van
  Maaren}}, \ and\ \bibinfo {author} {\bibfnamefont {C.}~\bibnamefont
  {Caleman}},\ }\href {\doibase 10.1093/bioinformatics/bts020} {\bibfield
  {journal} {\bibinfo  {journal} {Bioinformatics}\ }\textbf {\bibinfo {volume}
  {28}},\ \bibinfo {pages} {752} (\bibinfo {year} {2012})}\BibitemShut
  {NoStop}%
\bibitem [{\citenamefont {Mayoral}\ and\ \citenamefont
  {Goicochea}(2013)}]{RN31}%
  \BibitemOpen
  \bibfield  {author} {\bibinfo {author} {\bibfnamefont {E.}~\bibnamefont
  {Mayoral}}\ and\ \bibinfo {author} {\bibfnamefont {A.~G.}\ \bibnamefont
  {Goicochea}},\ }\href {\doibase 10.1063/1.4793742} {\bibfield  {journal}
  {\bibinfo  {journal} {J. Chem. Phys.}\ }\textbf {\bibinfo {volume} {138}},\
  \bibinfo {pages} {7} (\bibinfo {year} {2013})}\BibitemShut {NoStop}%
\bibitem [{\citenamefont {Yesylevskyy}\ \emph {et~al.}(2010)\citenamefont
  {Yesylevskyy}, \citenamefont {Schafer}, \citenamefont {Sengupta},\ and\
  \citenamefont {Marrink}}]{RN32}%
  \BibitemOpen
  \bibfield  {author} {\bibinfo {author} {\bibfnamefont {S.~O.}\ \bibnamefont
  {Yesylevskyy}}, \bibinfo {author} {\bibfnamefont {L.~V.}\ \bibnamefont
  {Schafer}}, \bibinfo {author} {\bibfnamefont {D.}~\bibnamefont {Sengupta}}, \
  and\ \bibinfo {author} {\bibfnamefont {S.~J.}\ \bibnamefont {Marrink}},\
  }\href {\doibase 10.1371/journal.pcbi.1000810} {\bibfield  {journal}
  {\bibinfo  {journal} {PPLOS Comput. Biol.}\ }\textbf {\bibinfo {volume} {6}}
  (\bibinfo {year} {2010}),\ 10.1371/journal.pcbi.1000810}\BibitemShut
  {NoStop}%
\bibitem [{\citenamefont {Lobanova}\ \emph {et~al.}(2015)\citenamefont
  {Lobanova}, \citenamefont {Avendano}, \citenamefont {Lafitte}, \citenamefont
  {Muller},\ and\ \citenamefont {Jackson}}]{RN33}%
  \BibitemOpen
  \bibfield  {author} {\bibinfo {author} {\bibfnamefont {O.}~\bibnamefont
  {Lobanova}}, \bibinfo {author} {\bibfnamefont {C.}~\bibnamefont {Avendano}},
  \bibinfo {author} {\bibfnamefont {T.}~\bibnamefont {Lafitte}}, \bibinfo
  {author} {\bibfnamefont {E.~A.}\ \bibnamefont {Muller}}, \ and\ \bibinfo
  {author} {\bibfnamefont {G.}~\bibnamefont {Jackson}},\ }\href {\doibase
  10.1080/00268976.2015.1004804} {\bibfield  {journal} {\bibinfo  {journal}
  {Mol. Phys.}\ }\textbf {\bibinfo {volume} {113}},\ \bibinfo {pages} {1228}
  (\bibinfo {year} {2015})}\BibitemShut {NoStop}%
\bibitem [{\citenamefont {Lobanova}\ \emph {et~al.}(2016)\citenamefont
  {Lobanova}, \citenamefont {Mejia}, \citenamefont {Jackson},\ and\
  \citenamefont {Muller}}]{RN34}%
  \BibitemOpen
  \bibfield  {author} {\bibinfo {author} {\bibfnamefont {O.}~\bibnamefont
  {Lobanova}}, \bibinfo {author} {\bibfnamefont {A.}~\bibnamefont {Mejia}},
  \bibinfo {author} {\bibfnamefont {G.}~\bibnamefont {Jackson}}, \ and\
  \bibinfo {author} {\bibfnamefont {E.~A.}\ \bibnamefont {Muller}},\ }\href
  {\doibase https://doi.org/10.1016/j.jct.2015.10.011} {\bibfield  {journal}
  {\bibinfo  {journal} {J. Chem. Thermodyn.}\ }\textbf {\bibinfo {volume}
  {93}},\ \bibinfo {pages} {320} (\bibinfo {year} {2016})}\BibitemShut
  {NoStop}%
\bibitem [{\citenamefont {Shinoda}, \citenamefont {Devane},\ and\ \citenamefont
  {Klein}(2007)}]{RN23}%
  \BibitemOpen
  \bibfield  {author} {\bibinfo {author} {\bibfnamefont {W.}~\bibnamefont
  {Shinoda}}, \bibinfo {author} {\bibfnamefont {R.}~\bibnamefont {Devane}}, \
  and\ \bibinfo {author} {\bibfnamefont {M.~L.}\ \bibnamefont {Klein}},\ }\href
  {\doibase 10.1080/08927020601054050} {\bibfield  {journal} {\bibinfo
  {journal} {Mol. Simulat.}\ }\textbf {\bibinfo {volume} {33}},\ \bibinfo
  {pages} {27} (\bibinfo {year} {2007})}\BibitemShut {NoStop}%
\bibitem [{\citenamefont {Marrink}\ \emph {et~al.}(2007)\citenamefont
  {Marrink}, \citenamefont {Risselada}, \citenamefont {Yefimov}, \citenamefont
  {Tieleman},\ and\ \citenamefont {de~Vries}}]{RN25}%
  \BibitemOpen
  \bibfield  {author} {\bibinfo {author} {\bibfnamefont {S.~J.}\ \bibnamefont
  {Marrink}}, \bibinfo {author} {\bibfnamefont {H.~J.}\ \bibnamefont
  {Risselada}}, \bibinfo {author} {\bibfnamefont {S.}~\bibnamefont {Yefimov}},
  \bibinfo {author} {\bibfnamefont {D.~P.}\ \bibnamefont {Tieleman}}, \ and\
  \bibinfo {author} {\bibfnamefont {A.~H.}\ \bibnamefont {de~Vries}},\ }\href
  {\doibase 10.1021/jp071097f} {\bibfield  {journal} {\bibinfo  {journal} {J.
  Phys. Chem. B}\ }\textbf {\bibinfo {volume} {111}},\ \bibinfo {pages} {7812}
  (\bibinfo {year} {2007})}\BibitemShut {NoStop}%
\bibitem [{\citenamefont {Monticelli}\ \emph {et~al.}(2008)\citenamefont
  {Monticelli}, \citenamefont {Kandasamy}, \citenamefont {Periole},
  \citenamefont {Larson}, \citenamefont {Tieleman},\ and\ \citenamefont
  {Marrink}}]{RN26}%
  \BibitemOpen
  \bibfield  {author} {\bibinfo {author} {\bibfnamefont {L.}~\bibnamefont
  {Monticelli}}, \bibinfo {author} {\bibfnamefont {S.~K.}\ \bibnamefont
  {Kandasamy}}, \bibinfo {author} {\bibfnamefont {X.}~\bibnamefont {Periole}},
  \bibinfo {author} {\bibfnamefont {R.~G.}\ \bibnamefont {Larson}}, \bibinfo
  {author} {\bibfnamefont {D.~P.}\ \bibnamefont {Tieleman}}, \ and\ \bibinfo
  {author} {\bibfnamefont {S.~J.}\ \bibnamefont {Marrink}},\ }\href {\doibase
  10.1021/ct700324x} {\bibfield  {journal} {\bibinfo  {journal} {J. Chem.
  Theory. Comput.}\ }\textbf {\bibinfo {volume} {4}},\ \bibinfo {pages} {819}
  (\bibinfo {year} {2008})}\BibitemShut {NoStop}%
\bibitem [{\citenamefont {Ghoufi}, \citenamefont {Malfreyt},\ and\
  \citenamefont {Tildesley}(2016)}]{RN27}%
  \BibitemOpen
  \bibfield  {author} {\bibinfo {author} {\bibfnamefont {A.}~\bibnamefont
  {Ghoufi}}, \bibinfo {author} {\bibfnamefont {P.}~\bibnamefont {Malfreyt}}, \
  and\ \bibinfo {author} {\bibfnamefont {D.~J.}\ \bibnamefont {Tildesley}},\
  }\href {\doibase 10.1039/c5cs00736d} {\bibfield  {journal} {\bibinfo
  {journal} {Chem. Soc. Rev.}\ }\textbf {\bibinfo {volume} {45}},\ \bibinfo
  {pages} {1387} (\bibinfo {year} {2016})}\BibitemShut {NoStop}%
\bibitem [{\citenamefont {Zubillaga}\ \emph {et~al.}(2013)\citenamefont
  {Zubillaga}, \citenamefont {Labastida}, \citenamefont {Cruz}, \citenamefont
  {Martínez}, \citenamefont {Sánchez},\ and\ \citenamefont
  {Alejandre}}]{RN29}%
  \BibitemOpen
  \bibfield  {author} {\bibinfo {author} {\bibfnamefont {R.~A.}\ \bibnamefont
  {Zubillaga}}, \bibinfo {author} {\bibfnamefont {A.}~\bibnamefont
  {Labastida}}, \bibinfo {author} {\bibfnamefont {B.}~\bibnamefont {Cruz}},
  \bibinfo {author} {\bibfnamefont {J.~C.}\ \bibnamefont {Martínez}}, \bibinfo
  {author} {\bibfnamefont {E.}~\bibnamefont {Sánchez}}, \ and\ \bibinfo
  {author} {\bibfnamefont {J.}~\bibnamefont {Alejandre}},\ }\href {\doibase
  10.1021/ct300976t} {\bibfield  {journal} {\bibinfo  {journal} {J. Chem.
  Theory Comput.}\ }\textbf {\bibinfo {volume} {9}},\ \bibinfo {pages} {1611}
  (\bibinfo {year} {2013})}\BibitemShut {NoStop}%
\bibitem [{\citenamefont {John}\ and\ \citenamefont {Csanyi}(2017)}]{RN781}%
  \BibitemOpen
  \bibfield  {author} {\bibinfo {author} {\bibfnamefont {S.~T.}\ \bibnamefont
  {John}}\ and\ \bibinfo {author} {\bibfnamefont {G.}~\bibnamefont {Csanyi}},\
  }\href {\doibase 10.1021/acs.jpcb.7b09636} {\bibfield  {journal} {\bibinfo
  {journal} {J. Phys. Chem. B}\ }\textbf {\bibinfo {volume} {121}},\ \bibinfo
  {pages} {10934} (\bibinfo {year} {2017})}\BibitemShut {NoStop}%
\bibitem [{\citenamefont {Sidky}\ and\ \citenamefont {Whitmer}(2018)}]{RN783}%
  \BibitemOpen
  \bibfield  {author} {\bibinfo {author} {\bibfnamefont {H.}~\bibnamefont
  {Sidky}}\ and\ \bibinfo {author} {\bibfnamefont {J.~K.}\ \bibnamefont
  {Whitmer}},\ }\href {\doibase 10.1063/1.5018708} {\bibfield  {journal}
  {\bibinfo  {journal} {J. Chem. Phys.}\ }\textbf {\bibinfo {volume} {148}},\
  \bibinfo {pages} {104111} (\bibinfo {year} {2018})}\BibitemShut {NoStop}%
\bibitem [{\citenamefont {Chmiela}\ \emph {et~al.}(2018)\citenamefont
  {Chmiela}, \citenamefont {Sauceda}, \citenamefont {Müller},\ and\
  \citenamefont {Tkatchenko}}]{RN785}%
  \BibitemOpen
  \bibfield  {author} {\bibinfo {author} {\bibfnamefont {S.}~\bibnamefont
  {Chmiela}}, \bibinfo {author} {\bibfnamefont {H.~E.}\ \bibnamefont
  {Sauceda}}, \bibinfo {author} {\bibfnamefont {K.-R.}\ \bibnamefont
  {Müller}}, \ and\ \bibinfo {author} {\bibfnamefont {A.}~\bibnamefont
  {Tkatchenko}},\ }\href {\doibase 10.1038/s41467-018-06169-2} {\bibfield
  {journal} {\bibinfo  {journal} {Nat. Commun.}\ }\textbf {\bibinfo {volume}
  {9}},\ \bibinfo {pages} {3887} (\bibinfo {year} {2018})}\BibitemShut
  {NoStop}%
\bibitem [{\citenamefont {Tang}, \citenamefont {Zhang},\ and\ \citenamefont
  {Karniadakis}(2018)}]{RN784}%
  \BibitemOpen
  \bibfield  {author} {\bibinfo {author} {\bibfnamefont {Y.-H.}\ \bibnamefont
  {Tang}}, \bibinfo {author} {\bibfnamefont {D.}~\bibnamefont {Zhang}}, \ and\
  \bibinfo {author} {\bibfnamefont {G.~E.}\ \bibnamefont {Karniadakis}},\
  }\href {\doibase 10.1063/1.5008630} {\bibfield  {journal} {\bibinfo
  {journal} {J. Chem. Phys.}\ }\textbf {\bibinfo {volume} {148}},\ \bibinfo
  {pages} {034101} (\bibinfo {year} {2018})}\BibitemShut {NoStop}%
\bibitem [{\citenamefont {Underwood}\ and\ \citenamefont
  {Greenwell}(2018)}]{RN36}%
  \BibitemOpen
  \bibfield  {author} {\bibinfo {author} {\bibfnamefont {T.~R.}\ \bibnamefont
  {Underwood}}\ and\ \bibinfo {author} {\bibfnamefont {H.~C.}\ \bibnamefont
  {Greenwell}},\ }\href {\doibase 10.1038/s41598-017-18633-y} {\bibfield
  {journal} {\bibinfo  {journal} {Sci. Rep.}\ }\textbf {\bibinfo {volume}
  {8}},\ \bibinfo {pages} {11} (\bibinfo {year} {2018})}\BibitemShut {NoStop}%
\bibitem [{\citenamefont {Abascal}\ and\ \citenamefont {Vega}(2005)}]{RN37}%
  \BibitemOpen
  \bibfield  {author} {\bibinfo {author} {\bibfnamefont {J.~L.~F.}\
  \bibnamefont {Abascal}}\ and\ \bibinfo {author} {\bibfnamefont
  {C.}~\bibnamefont {Vega}},\ }\href {\doibase 10.1063/1.2121687} {\bibfield
  {journal} {\bibinfo  {journal} {J. Chem. Phys.}\ }\textbf {\bibinfo {volume}
  {123}},\ \bibinfo {pages} {234505} (\bibinfo {year} {2005})}\BibitemShut
  {NoStop}%
\bibitem [{\citenamefont {Martin}\ and\ \citenamefont {Siepmann}(1998)}]{RN38}%
  \BibitemOpen
  \bibfield  {author} {\bibinfo {author} {\bibfnamefont {M.~G.}\ \bibnamefont
  {Martin}}\ and\ \bibinfo {author} {\bibfnamefont {J.~I.}\ \bibnamefont
  {Siepmann}},\ }\href {\doibase 10.1021/jp972543+} {\bibfield  {journal}
  {\bibinfo  {journal} {J. Phys. Chem. B}\ }\textbf {\bibinfo {volume} {102}},\
  \bibinfo {pages} {2569} (\bibinfo {year} {1998})}\BibitemShut {NoStop}%
\bibitem [{\citenamefont {Neyt}\ \emph {et~al.}(2014)\citenamefont {Neyt},
  \citenamefont {Wender}, \citenamefont {Lachet}, \citenamefont {Ghoufi},\ and\
  \citenamefont {Malfreyt}}]{RN39}%
  \BibitemOpen
  \bibfield  {author} {\bibinfo {author} {\bibfnamefont {J.~C.}\ \bibnamefont
  {Neyt}}, \bibinfo {author} {\bibfnamefont {A.}~\bibnamefont {Wender}},
  \bibinfo {author} {\bibfnamefont {V.}~\bibnamefont {Lachet}}, \bibinfo
  {author} {\bibfnamefont {A.}~\bibnamefont {Ghoufi}}, \ and\ \bibinfo {author}
  {\bibfnamefont {P.}~\bibnamefont {Malfreyt}},\ }\href {\doibase
  10.1021/ct500053c} {\bibfield  {journal} {\bibinfo  {journal} {J. Chem.
  Theory. Comput.}\ }\textbf {\bibinfo {volume} {10}},\ \bibinfo {pages} {1887}
  (\bibinfo {year} {2014})}\BibitemShut {NoStop}%
\bibitem [{\citenamefont {Ashbaugh}, \citenamefont {Liu},\ and\ \citenamefont
  {Surampudi}(2011)}]{RN40}%
  \BibitemOpen
  \bibfield  {author} {\bibinfo {author} {\bibfnamefont {H.~S.}\ \bibnamefont
  {Ashbaugh}}, \bibinfo {author} {\bibfnamefont {L.}~\bibnamefont {Liu}}, \
  and\ \bibinfo {author} {\bibfnamefont {L.~N.}\ \bibnamefont {Surampudi}},\
  }\href {\doibase 10.1063/1.3623267} {\bibfield  {journal} {\bibinfo
  {journal} {J. Chem. Phys.}\ }\textbf {\bibinfo {volume} {135}},\ \bibinfo
  {pages} {054510} (\bibinfo {year} {2011})}\BibitemShut {NoStop}%
\bibitem [{\citenamefont {Kaminski}\ and\ \citenamefont
  {Jorgensen}(1996)}]{RN41}%
  \BibitemOpen
  \bibfield  {author} {\bibinfo {author} {\bibfnamefont {G.}~\bibnamefont
  {Kaminski}}\ and\ \bibinfo {author} {\bibfnamefont {W.~L.}\ \bibnamefont
  {Jorgensen}},\ }\href {\doibase 10.1021/jp9624257} {\bibfield  {journal}
  {\bibinfo  {journal} {J. Phys. Chem.}\ }\textbf {\bibinfo {volume} {100}},\
  \bibinfo {pages} {18010} (\bibinfo {year} {1996})}\BibitemShut {NoStop}%
\bibitem [{\citenamefont {Shi}\ and\ \citenamefont {Guo}(2010)}]{RN724}%
  \BibitemOpen
  \bibfield  {author} {\bibinfo {author} {\bibfnamefont {W.-X.}\ \bibnamefont
  {Shi}}\ and\ \bibinfo {author} {\bibfnamefont {H.-X.}\ \bibnamefont {Guo}},\
  }\href {\doibase 10.1021/jp100868p} {\bibfield  {journal} {\bibinfo
  {journal} {J. Phys. Chem. B}\ }\textbf {\bibinfo {volume} {114}},\ \bibinfo
  {pages} {6365} (\bibinfo {year} {2010})}\BibitemShut {NoStop}%
\bibitem [{\citenamefont {Hess}\ \emph {et~al.}(1997)\citenamefont {Hess},
  \citenamefont {Bekker}, \citenamefont {Berendsen},\ and\ \citenamefont
  {Fraaije}}]{RN42}%
  \BibitemOpen
  \bibfield  {author} {\bibinfo {author} {\bibfnamefont {B.}~\bibnamefont
  {Hess}}, \bibinfo {author} {\bibfnamefont {H.}~\bibnamefont {Bekker}},
  \bibinfo {author} {\bibfnamefont {H.~J.~C.}\ \bibnamefont {Berendsen}}, \
  and\ \bibinfo {author} {\bibfnamefont {J.~G. E.~M.}\ \bibnamefont
  {Fraaije}},\ }\href {\doibase
  10.1002/(SICI)1096-987X(199709)18:12<1463::AID-JCC4>3.0.CO;2-H} {\bibfield
  {journal} {\bibinfo  {journal} {J. Comput. Chem.}\ }\textbf {\bibinfo
  {volume} {18}},\ \bibinfo {pages} {1463} (\bibinfo {year}
  {1997})}\BibitemShut {NoStop}%
\bibitem [{\citenamefont {Biscay}\ \emph {et~al.}(2009)\citenamefont {Biscay},
  \citenamefont {Ghoufi}, \citenamefont {Goujon}, \citenamefont {Lachet},\ and\
  \citenamefont {Malfreyt}}]{RN53}%
  \BibitemOpen
  \bibfield  {author} {\bibinfo {author} {\bibfnamefont {F.}~\bibnamefont
  {Biscay}}, \bibinfo {author} {\bibfnamefont {A.}~\bibnamefont {Ghoufi}},
  \bibinfo {author} {\bibfnamefont {F.}~\bibnamefont {Goujon}}, \bibinfo
  {author} {\bibfnamefont {V.}~\bibnamefont {Lachet}}, \ and\ \bibinfo {author}
  {\bibfnamefont {P.}~\bibnamefont {Malfreyt}},\ }\href {\doibase
  10.1063/1.3132708} {\bibfield  {journal} {\bibinfo  {journal} {J. Chem.
  Phys.}\ }\textbf {\bibinfo {volume} {130}},\ \bibinfo {pages} {184710}
  (\bibinfo {year} {2009})}\BibitemShut {NoStop}%
\bibitem [{\citenamefont {Mitrinović}\ \emph {et~al.}(2000)\citenamefont
  {Mitrinović}, \citenamefont {Tikhonov}, \citenamefont {Li}, \citenamefont
  {Huang},\ and\ \citenamefont {Schlossman}}]{RN612}%
  \BibitemOpen
  \bibfield  {author} {\bibinfo {author} {\bibfnamefont {D.~M.}\ \bibnamefont
  {Mitrinović}}, \bibinfo {author} {\bibfnamefont {A.~M.}\ \bibnamefont
  {Tikhonov}}, \bibinfo {author} {\bibfnamefont {M.}~\bibnamefont {Li}},
  \bibinfo {author} {\bibfnamefont {Z.}~\bibnamefont {Huang}}, \ and\ \bibinfo
  {author} {\bibfnamefont {M.~L.}\ \bibnamefont {Schlossman}},\ }\href
  {https://link.aps.org/doi/10.1103/PhysRevLett.85.582} {\bibfield  {journal}
  {\bibinfo  {journal} {Phys. Rev. Lett.}\ }\textbf {\bibinfo {volume} {85}},\
  \bibinfo {pages} {582} (\bibinfo {year} {2000})}\BibitemShut {NoStop}%
\bibitem [{\citenamefont {Hantal}\ \emph {et~al.}(2010)\citenamefont {Hantal},
  \citenamefont {Darvas}, \citenamefont {Partay}, \citenamefont {Horvai},\ and\
  \citenamefont {Jedlovszky}}]{RN54}%
  \BibitemOpen
  \bibfield  {author} {\bibinfo {author} {\bibfnamefont {G.}~\bibnamefont
  {Hantal}}, \bibinfo {author} {\bibfnamefont {M.}~\bibnamefont {Darvas}},
  \bibinfo {author} {\bibfnamefont {L.~B.}\ \bibnamefont {Partay}}, \bibinfo
  {author} {\bibfnamefont {G.}~\bibnamefont {Horvai}}, \ and\ \bibinfo {author}
  {\bibfnamefont {P.}~\bibnamefont {Jedlovszky}},\ }\href {\doibase
  10.1088/0953-8984/22/28/284112} {\bibfield  {journal} {\bibinfo  {journal}
  {J. Phys. Condens. Matter}\ }\textbf {\bibinfo {volume} {22}} (\bibinfo
  {year} {2010}),\ 10.1088/0953-8984/22/28/284112}\BibitemShut {NoStop}%
\bibitem [{\citenamefont {Bresme}\ \emph {et~al.}(2008)\citenamefont {Bresme},
  \citenamefont {Chacón}, \citenamefont {Tarazona},\ and\ \citenamefont
  {Tay}}]{RN55}%
  \BibitemOpen
  \bibfield  {author} {\bibinfo {author} {\bibfnamefont {F.}~\bibnamefont
  {Bresme}}, \bibinfo {author} {\bibfnamefont {E.}~\bibnamefont {Chacón}},
  \bibinfo {author} {\bibfnamefont {P.}~\bibnamefont {Tarazona}}, \ and\
  \bibinfo {author} {\bibfnamefont {K.}~\bibnamefont {Tay}},\ }\href
  {https://link.aps.org/doi/10.1103/PhysRevLett.101.056102} {\bibfield
  {journal} {\bibinfo  {journal} {Phys. Rev. Lett.}\ }\textbf {\bibinfo
  {volume} {101}},\ \bibinfo {pages} {056102} (\bibinfo {year}
  {2008})}\BibitemShut {NoStop}%
\bibitem [{\citenamefont {Bresme}, \citenamefont {Chacon},\ and\ \citenamefont
  {Tarazona}(2008)}]{RN56}%
  \BibitemOpen
  \bibfield  {author} {\bibinfo {author} {\bibfnamefont {F.}~\bibnamefont
  {Bresme}}, \bibinfo {author} {\bibfnamefont {E.}~\bibnamefont {Chacon}}, \
  and\ \bibinfo {author} {\bibfnamefont {P.}~\bibnamefont {Tarazona}},\ }\href
  {\doibase 10.1039/B807437M} {\bibfield  {journal} {\bibinfo  {journal} {Phys.
  Chem. Chem. Phys.}\ }\textbf {\bibinfo {volume} {10}},\ \bibinfo {pages}
  {4704} (\bibinfo {year} {2008})}\BibitemShut {NoStop}%
\bibitem [{\citenamefont {Partay~Livia}\ \emph {et~al.}(2007)\citenamefont
  {Partay~Livia}, \citenamefont {Hantal}, \citenamefont {Jedlovszky},
  \citenamefont {Vincze},\ and\ \citenamefont {Horvai}}]{RN57}%
  \BibitemOpen
  \bibfield  {author} {\bibinfo {author} {\bibfnamefont {B.}~\bibnamefont
  {Partay~Livia}}, \bibinfo {author} {\bibfnamefont {G.}~\bibnamefont
  {Hantal}}, \bibinfo {author} {\bibfnamefont {P.}~\bibnamefont {Jedlovszky}},
  \bibinfo {author} {\bibfnamefont {r.}~\bibnamefont {Vincze}}, \ and\ \bibinfo
  {author} {\bibfnamefont {G.}~\bibnamefont {Horvai}},\ }\href {\doibase
  10.1002/jcc.20852} {\bibfield  {journal} {\bibinfo  {journal} {J. Comput.
  Chem.}\ }\textbf {\bibinfo {volume} {29}},\ \bibinfo {pages} {945} (\bibinfo
  {year} {2007})}\BibitemShut {NoStop}%
\bibitem [{\citenamefont {Bresme}, \citenamefont {Chacón},\ and\ \citenamefont
  {Tarazona}(2010)}]{RN72}%
  \BibitemOpen
  \bibfield  {author} {\bibinfo {author} {\bibfnamefont {F.}~\bibnamefont
  {Bresme}}, \bibinfo {author} {\bibfnamefont {E.}~\bibnamefont {Chacón}}, \
  and\ \bibinfo {author} {\bibfnamefont {P.}~\bibnamefont {Tarazona}},\ }\href
  {\doibase 10.1080/00268976.2010.496376} {\bibfield  {journal} {\bibinfo
  {journal} {Mol. Phys.}\ }\textbf {\bibinfo {volume} {108}},\ \bibinfo {pages}
  {1887} (\bibinfo {year} {2010})}\BibitemShut {NoStop}%
\bibitem [{\citenamefont {Roddy}\ and\ \citenamefont {Coleman}(1968)}]{RN956}%
  \BibitemOpen
  \bibfield  {author} {\bibinfo {author} {\bibfnamefont {J.~W.}\ \bibnamefont
  {Roddy}}\ and\ \bibinfo {author} {\bibfnamefont {C.~F.}\ \bibnamefont
  {Coleman}},\ }\href {\doibase https://doi.org/10.1016/0039-9140(68)80050-5}
  {\bibfield  {journal} {\bibinfo  {journal} {Talanta}\ }\textbf {\bibinfo
  {volume} {15}},\ \bibinfo {pages} {1281} (\bibinfo {year}
  {1968})}\BibitemShut {NoStop}%
\bibitem [{\citenamefont {Nicolas}\ and\ \citenamefont {Smit}(2002)}]{RN73}%
  \BibitemOpen
  \bibfield  {author} {\bibinfo {author} {\bibfnamefont {J.~P.}\ \bibnamefont
  {Nicolas}}\ and\ \bibinfo {author} {\bibfnamefont {B.}~\bibnamefont {Smit}},\
  }\href {\doibase 10.1080/00268970210130182} {\bibfield  {journal} {\bibinfo
  {journal} {Mol. Phys.}\ }\textbf {\bibinfo {volume} {100}},\ \bibinfo {pages}
  {2471} (\bibinfo {year} {2002})}\BibitemShut {NoStop}%
\bibitem [{\citenamefont {Irving}\ and\ \citenamefont {Kirkwood}(1950)}]{RN62}%
  \BibitemOpen
  \bibfield  {author} {\bibinfo {author} {\bibfnamefont {J.~H.}\ \bibnamefont
  {Irving}}\ and\ \bibinfo {author} {\bibfnamefont {J.~G.}\ \bibnamefont
  {Kirkwood}},\ }\href {\doibase 10.1063/1.1747782} {\bibfield  {journal}
  {\bibinfo  {journal} {J. Chem. Phys.}\ }\textbf {\bibinfo {volume} {18}},\
  \bibinfo {pages} {817} (\bibinfo {year} {1950})}\BibitemShut {NoStop}%
\bibitem [{\citenamefont {Ghoufi}\ \emph {et~al.}(2008)\citenamefont {Ghoufi},
  \citenamefont {Goujon}, \citenamefont {Lachet},\ and\ \citenamefont
  {Malfreyt}}]{RN63}%
  \BibitemOpen
  \bibfield  {author} {\bibinfo {author} {\bibfnamefont {A.}~\bibnamefont
  {Ghoufi}}, \bibinfo {author} {\bibfnamefont {F.}~\bibnamefont {Goujon}},
  \bibinfo {author} {\bibfnamefont {V.}~\bibnamefont {Lachet}}, \ and\ \bibinfo
  {author} {\bibfnamefont {P.}~\bibnamefont {Malfreyt}},\ }\href
  {https://link.aps.org/doi/10.1103/PhysRevE.77.031601} {\bibfield  {journal}
  {\bibinfo  {journal} {Phys. Rev. E.}\ }\textbf {\bibinfo {volume} {77}},\
  \bibinfo {pages} {031601} (\bibinfo {year} {2008})}\BibitemShut {NoStop}%
\bibitem [{\citenamefont {Lovett}\ and\ \citenamefont {Baus}(1997)}]{RN66}%
  \BibitemOpen
  \bibfield  {author} {\bibinfo {author} {\bibfnamefont {R.}~\bibnamefont
  {Lovett}}\ and\ \bibinfo {author} {\bibfnamefont {M.}~\bibnamefont {Baus}},\
  }\href {\doibase 10.1063/1.473384} {\bibfield  {journal} {\bibinfo  {journal}
  {J. Chem. Phys.}\ }\textbf {\bibinfo {volume} {106}},\ \bibinfo {pages} {635}
  (\bibinfo {year} {1997})}\BibitemShut {NoStop}%
\bibitem [{\citenamefont {Vanegas}, \citenamefont {Torres-Sánchez},\ and\
  \citenamefont {Arroyo}(2014)}]{RN67}%
  \BibitemOpen
  \bibfield  {author} {\bibinfo {author} {\bibfnamefont {J.~M.}\ \bibnamefont
  {Vanegas}}, \bibinfo {author} {\bibfnamefont {A.}~\bibnamefont
  {Torres-Sánchez}}, \ and\ \bibinfo {author} {\bibfnamefont {M.}~\bibnamefont
  {Arroyo}},\ }\href {\doibase 10.1021/ct4008926} {\bibfield  {journal}
  {\bibinfo  {journal} {J. Chem. Theory Comput.}\ }\textbf {\bibinfo {volume}
  {10}},\ \bibinfo {pages} {691} (\bibinfo {year} {2014})}\BibitemShut
  {NoStop}%
\bibitem [{\citenamefont {Admal}\ and\ \citenamefont {Tadmor}(2010)}]{RN69}%
  \BibitemOpen
  \bibfield  {author} {\bibinfo {author} {\bibfnamefont {N.~C.}\ \bibnamefont
  {Admal}}\ and\ \bibinfo {author} {\bibfnamefont {E.~B.}\ \bibnamefont
  {Tadmor}},\ }\href {\doibase 10.1007/s10659-010-9249-6} {\bibfield  {journal}
  {\bibinfo  {journal} {J. Elasticity.}\ }\textbf {\bibinfo {volume} {100}},\
  \bibinfo {pages} {63} (\bibinfo {year} {2010})}\BibitemShut {NoStop}%
\bibitem [{\citenamefont {Nicolas}\ and\ \citenamefont
  {de~Souza}(2004)}]{RN74}%
  \BibitemOpen
  \bibfield  {author} {\bibinfo {author} {\bibfnamefont {J.~P.}\ \bibnamefont
  {Nicolas}}\ and\ \bibinfo {author} {\bibfnamefont {N.~R.}\ \bibnamefont
  {de~Souza}},\ }\href {\doibase 10.1063/1.1629278} {\bibfield  {journal}
  {\bibinfo  {journal} {J. Chem. Phys.}\ }\textbf {\bibinfo {volume} {120}},\
  \bibinfo {pages} {2464} (\bibinfo {year} {2004})}\BibitemShut {NoStop}%
\bibitem [{\citenamefont {Zeppieri}, \citenamefont {Rodríguez},\ and\
  \citenamefont {López~de Ramos}(2001)}]{RN922}%
  \BibitemOpen
  \bibfield  {author} {\bibinfo {author} {\bibfnamefont {S.}~\bibnamefont
  {Zeppieri}}, \bibinfo {author} {\bibfnamefont {J.}~\bibnamefont
  {Rodríguez}}, \ and\ \bibinfo {author} {\bibfnamefont {A.~L.}\ \bibnamefont
  {López~de Ramos}},\ }\href {\doibase 10.1021/je000245r} {\bibfield
  {journal} {\bibinfo  {journal} {J. Chem. Eng. Data}\ }\textbf {\bibinfo
  {volume} {46}},\ \bibinfo {pages} {1086} (\bibinfo {year}
  {2001})}\BibitemShut {NoStop}%
\bibitem [{\citenamefont {Takahashi}\ and\ \citenamefont {Morita}(2013)}]{RN8}%
  \BibitemOpen
  \bibfield  {author} {\bibinfo {author} {\bibfnamefont {H.}~\bibnamefont
  {Takahashi}}\ and\ \bibinfo {author} {\bibfnamefont {A.}~\bibnamefont
  {Morita}},\ }\href {\doibase 10.1016/j.cplett.2013.04.041} {\bibfield
  {journal} {\bibinfo  {journal} {Chem. Phys. Lett.}\ }\textbf {\bibinfo
  {volume} {573}},\ \bibinfo {pages} {35} (\bibinfo {year} {2013})}\BibitemShut
  {NoStop}%
\bibitem [{\citenamefont {Ghoufi}\ and\ \citenamefont {Malfreyt}(2011)}]{RN10}%
  \BibitemOpen
  \bibfield  {author} {\bibinfo {author} {\bibfnamefont {A.}~\bibnamefont
  {Ghoufi}}\ and\ \bibinfo {author} {\bibfnamefont {P.}~\bibnamefont
  {Malfreyt}},\ }\href {\doibase 10.1063/1.3632991} {\bibfield  {journal}
  {\bibinfo  {journal} {J. Chem. Phys.}\ }\textbf {\bibinfo {volume} {135}},\
  \bibinfo {pages} {104105} (\bibinfo {year} {2011})}\BibitemShut {NoStop}%
\bibitem [{\citenamefont {Malijevsky}\ and\ \citenamefont
  {Jackson}(2012)}]{RN12}%
  \BibitemOpen
  \bibfield  {author} {\bibinfo {author} {\bibfnamefont {A.}~\bibnamefont
  {Malijevsky}}\ and\ \bibinfo {author} {\bibfnamefont {G.}~\bibnamefont
  {Jackson}},\ }\href {\doibase 10.1088/0953-8984/24/46/464121} {\bibfield
  {journal} {\bibinfo  {journal} {J. Phys. Condens. Matter}\ }\textbf {\bibinfo
  {volume} {24}},\ \bibinfo {pages} {464121} (\bibinfo {year}
  {2012})}\BibitemShut {NoStop}%
\bibitem [{\citenamefont {Xiu}\ and\ \citenamefont {Karniadakis}(2002)}]{RN77}%
  \BibitemOpen
  \bibfield  {author} {\bibinfo {author} {\bibfnamefont {D.}~\bibnamefont
  {Xiu}}\ and\ \bibinfo {author} {\bibfnamefont {G.}~\bibnamefont
  {Karniadakis}},\ }\href {\doibase 10.1137/S1064827501387826} {\bibfield
  {journal} {\bibinfo  {journal} {SIAM J. Sci. Comput.}\ }\textbf {\bibinfo
  {volume} {24}},\ \bibinfo {pages} {619} (\bibinfo {year} {2002})}\BibitemShut
  {NoStop}%
\bibitem [{\citenamefont {Lei}\ \emph {et~al.}(2015)\citenamefont {Lei},
  \citenamefont {Yang}, \citenamefont {Zheng}, \citenamefont {Lin},\ and\
  \citenamefont {Baker}}]{RN75}%
  \BibitemOpen
  \bibfield  {author} {\bibinfo {author} {\bibfnamefont {H.}~\bibnamefont
  {Lei}}, \bibinfo {author} {\bibfnamefont {X.}~\bibnamefont {Yang}}, \bibinfo
  {author} {\bibfnamefont {B.}~\bibnamefont {Zheng}}, \bibinfo {author}
  {\bibfnamefont {G.}~\bibnamefont {Lin}}, \ and\ \bibinfo {author}
  {\bibfnamefont {N.~A.}\ \bibnamefont {Baker}},\ }\href {\doibase
  10.1137/140981587} {\bibfield  {journal} {\bibinfo  {journal} {Multiscale
  Model Sim.}\ }\textbf {\bibinfo {volume} {13}},\ \bibinfo {pages} {1327}
  (\bibinfo {year} {2015})}\BibitemShut {NoStop}%
\bibitem [{\citenamefont {Li}\ \emph {et~al.}(2016)\citenamefont {Li},
  \citenamefont {Bian}, \citenamefont {Yang},\ and\ \citenamefont
  {Karniadakis}}]{RN76}%
  \BibitemOpen
  \bibfield  {author} {\bibinfo {author} {\bibfnamefont {Z.}~\bibnamefont
  {Li}}, \bibinfo {author} {\bibfnamefont {X.}~\bibnamefont {Bian}}, \bibinfo
  {author} {\bibfnamefont {X.}~\bibnamefont {Yang}}, \ and\ \bibinfo {author}
  {\bibfnamefont {G.~E.}\ \bibnamefont {Karniadakis}},\ }\href {\doibase
  10.1063/1.4959121} {\bibfield  {journal} {\bibinfo  {journal} {J. Chem.
  Phys.}\ }\textbf {\bibinfo {volume} {145}},\ \bibinfo {pages} {044102}
  (\bibinfo {year} {2016})}\BibitemShut {NoStop}%
\bibitem [{\citenamefont {Gerstner}\ and\ \citenamefont
  {Griebel}(1998)}]{RN78}%
  \BibitemOpen
  \bibfield  {author} {\bibinfo {author} {\bibfnamefont {T.}~\bibnamefont
  {Gerstner}}\ and\ \bibinfo {author} {\bibfnamefont {M.}~\bibnamefont
  {Griebel}},\ }\href {\doibase 10.1023/a:1019129717644} {\bibfield  {journal}
  {\bibinfo  {journal} {Numer. Algorithms.}\ }\textbf {\bibinfo {volume}
  {18}},\ \bibinfo {pages} {209} (\bibinfo {year} {1998})}\BibitemShut
  {NoStop}%
\bibitem [{\citenamefont {Xiu}\ and\ \citenamefont {Hesthaven}(2005)}]{RN79}%
  \BibitemOpen
  \bibfield  {author} {\bibinfo {author} {\bibfnamefont {D.}~\bibnamefont
  {Xiu}}\ and\ \bibinfo {author} {\bibfnamefont {J.}~\bibnamefont
  {Hesthaven}},\ }\href {\doibase 10.1137/040615201} {\bibfield  {journal}
  {\bibinfo  {journal} {SIAM J. Sci. Comput.}\ }\textbf {\bibinfo {volume}
  {27}},\ \bibinfo {pages} {1118} (\bibinfo {year} {2005})}\BibitemShut
  {NoStop}%
\bibitem [{\citenamefont {Stone}(1974)}]{RN303}%
  \BibitemOpen
  \bibfield  {author} {\bibinfo {author} {\bibfnamefont {M.}~\bibnamefont
  {Stone}},\ }\href {http://www.jstor.org/stable/2984809} {\bibfield  {journal}
  {\bibinfo  {journal} {J. R. Stat. Soc. Series. B.}\ }\textbf {\bibinfo
  {volume} {36}},\ \bibinfo {pages} {111} (\bibinfo {year} {1974})}\BibitemShut
  {NoStop}%
\bibitem [{\citenamefont {Sega}\ \emph {et~al.}(2016)\citenamefont {Sega},
  \citenamefont {Fábián}, \citenamefont {Horvai},\ and\ \citenamefont
  {Jedlovszky}}]{RN58}%
  \BibitemOpen
  \bibfield  {author} {\bibinfo {author} {\bibfnamefont {M.}~\bibnamefont
  {Sega}}, \bibinfo {author} {\bibfnamefont {B.}~\bibnamefont {Fábián}},
  \bibinfo {author} {\bibfnamefont {G.}~\bibnamefont {Horvai}}, \ and\ \bibinfo
  {author} {\bibfnamefont {P.}~\bibnamefont {Jedlovszky}},\ }\href {\doibase
  10.1021/acs.jpcc.6b09880} {\bibfield  {journal} {\bibinfo  {journal} {J.
  Phys. Chem. C}\ }\textbf {\bibinfo {volume} {120}},\ \bibinfo {pages} {27468}
  (\bibinfo {year} {2016})}\BibitemShut {NoStop}%
\bibitem [{\citenamefont {Sega}\ \emph {et~al.}(2013)\citenamefont {Sega},
  \citenamefont {Kantorovich}, \citenamefont {Jedlovszky},\ and\ \citenamefont
  {Jorge}}]{RN59}%
  \BibitemOpen
  \bibfield  {author} {\bibinfo {author} {\bibfnamefont {M.}~\bibnamefont
  {Sega}}, \bibinfo {author} {\bibfnamefont {S.~S.}\ \bibnamefont
  {Kantorovich}}, \bibinfo {author} {\bibfnamefont {P.}~\bibnamefont
  {Jedlovszky}}, \ and\ \bibinfo {author} {\bibfnamefont {M.}~\bibnamefont
  {Jorge}},\ }\href {\doibase 10.1063/1.4776196} {\bibfield  {journal}
  {\bibinfo  {journal} {J. Chem. Phys.}\ }\textbf {\bibinfo {volume} {138}},\
  \bibinfo {pages} {044110} (\bibinfo {year} {2013})}\BibitemShut {NoStop}%
\bibitem [{\citenamefont {Sega}\ \emph {et~al.}(2018)\citenamefont {Sega},
  \citenamefont {Hantal}, \citenamefont {Fábián},\ and\ \citenamefont
  {Jedlovszky}}]{RN709}%
  \BibitemOpen
  \bibfield  {author} {\bibinfo {author} {\bibfnamefont {M.}~\bibnamefont
  {Sega}}, \bibinfo {author} {\bibfnamefont {G.}~\bibnamefont {Hantal}},
  \bibinfo {author} {\bibfnamefont {B.}~\bibnamefont {Fábián}}, \ and\
  \bibinfo {author} {\bibfnamefont {P.}~\bibnamefont {Jedlovszky}},\ }\href
  {\doibase 10.1002/jcc.25384} {\bibfield  {journal} {\bibinfo  {journal} {J.
  Comput. Chem.}\ }\textbf {\bibinfo {volume} {39}},\ \bibinfo {pages} {2118}
  (\bibinfo {year} {2018})}\BibitemShut {NoStop}%
\bibitem [{\citenamefont {Ollila}\ \emph {et~al.}(2009)\citenamefont {Ollila},
  \citenamefont {Risselada}, \citenamefont {Louhivuori}, \citenamefont
  {Lindahl}, \citenamefont {Vattulainen},\ and\ \citenamefont
  {Marrink}}]{RN245}%
  \BibitemOpen
  \bibfield  {author} {\bibinfo {author} {\bibfnamefont {O.~H.~S.}\
  \bibnamefont {Ollila}}, \bibinfo {author} {\bibfnamefont {H.~J.}\
  \bibnamefont {Risselada}}, \bibinfo {author} {\bibfnamefont {M.}~\bibnamefont
  {Louhivuori}}, \bibinfo {author} {\bibfnamefont {E.}~\bibnamefont {Lindahl}},
  \bibinfo {author} {\bibfnamefont {I.}~\bibnamefont {Vattulainen}}, \ and\
  \bibinfo {author} {\bibfnamefont {S.~J.}\ \bibnamefont {Marrink}},\ }\href
  {https://link.aps.org/doi/10.1103/PhysRevLett.102.078101} {\bibfield
  {journal} {\bibinfo  {journal} {Phys. Rev. Lett.}\ }\textbf {\bibinfo
  {volume} {102}},\ \bibinfo {pages} {078101} (\bibinfo {year}
  {2009})}\BibitemShut {NoStop}%
\end{thebibliography}%
\end{document}